\documentclass[smallextended]{svjour3}    
\smartqed 
\usepackage{graphicx}

\usepackage{lineno}

\usepackage{amsmath}
\usepackage{stackengine}
\usepackage[customcolors,shade]{hf-tikz}

\usepackage{amssymb}

\usepackage{setspace}

\usepackage{float}

\journalname{Theoretical Ecology}
\begin{document}

\onehalfspacing

\title{Interplay of spatial dynamics and local adaptation shapes species lifetime distributions and species-area relationships
\thanks{This work was supported by the German Research Foundation (DFG) under contract numbers Dr300/12 and Dr300/13. KTA was additionally supported by the French National Research Agency (ANR) through project ARSENIC (grant no. 14-CE02-0012). }
}

\titlerunning{Interplay of spatial dynamics and local adaptation}    

\author{Tobias Rogge \and
    David Jones \and
    Barbara Drossel \and
	Korinna T. Allhoff
}


\institute{T. Rogge \at
      Technische Universit\"{a}t Darmstadt, Institut f\"{u}r Festk\"{o}rperphysik, Darmstadt, Germany. \\
      \email{rogge@fkp.tu-darmstadt.de}      
      \and
      David Jones \at
      Technische Universit\"{a}t Darmstadt, Institut f\"{u}r Festk\"{o}rperphysik, Darmstadt, Germany. \\
      \email{jones@fkp.tu-darmstadt.de} 
      \and
		  Barbara Drossel \at
      Technische Universit\"{a}t Darmstadt, Institut f\"{u}r Festk\"{o}rperphysik, Darmstadt, Germany. \\
      \email{drossel@fkp.tu-darmstadt.de}     
      \and
      Korinna T. Allhoff \at
      Technische Universit\"{a}t Darmstadt, Institut f\"{u}r Festk\"{o}rperphysik, Darmstadt, Germany; \\
      Universit\'{e} Pierre et Marie Curie, Institut d'\'{e}cologie et des sciences de l'environnement de Paris, Paris, France; \\
      \emph{Present address:} Eberhard Karls Universit{\"a}t T{\"u}bingen, Institut f{\"u}r Evolution und {\"O}kologie, T{\"u}bingen, Germany. \\
      ORCID: 0000-0003-0164-7618.\\
      \email{korinna.allhoff@uni-tuebingen.de} 
}

\date{Received: date / Accepted: date}

\maketitle


\clearpage
\begin{abstract}
The distributions of species lifetimes and species in space are related, since species with good local survival chances have more time to colonize new habitats and species inhabiting large areas have higher chances to survive local disturbances. Yet, both distributions have been discussed in mostly separate communities. Here, we study both patterns simultaneously using a spatially explicit, evolutionary meta-food web model, consisting of a grid of patches, where each patch contains a local food web. Species survival depends on predation and competition interactions, which in turn depend on species body masses as the key traits. The system evolves due to the migration of species to neighboring patches, the addition of new species as modifications of existing species, and local extinction events. The structure of each local food web thus emerges in a self-organized manner as the highly non-trivial outcome of the relative time scales of these processes. Our model generates a large variety of complex, multi-trophic networks and therefore serves as a powerful tool to investigate ecosystems on long temporal and large spatial scales. We find that the observed lifetime distributions and species-area relations resemble power laws over appropriately chosen parameter ranges and thus agree qualitatively with empirical findings. Moreover, we observe strong finite-size effects, and a dependence of the relationships on the trophic level of the species. By comparing our results to simple neutral models found in the literature, we identify the features that affect the values of the exponents.

\end{abstract}

\clearpage
\section{Introduction}
\bigskip

The history of life on earth is marked by five major mass extinction events, which were caused by catastrophic external triggers such as volcanism or meteorite impacts, and led each to the disappearance of more than three quarters of all species during geologically short time intervals \cite{Raup1982,Raup1986}. Currently, the world faces one of the largest extinction waves ever, which is thought to be man-made \cite{Barnosky2011}. But biological extinctions do not only happen in response to external drivers. They also occur due to intrinsic processes, such as the dynamic trophic and competitive interactions among species \cite{Binzer2011,Allhoff2015a}. The fossil record of marine animal families actually suggests a spectrum of extinction events at all scales, with periods of low extinction activity alternating with occasional pulses of much higher extinction rates \cite{Raup1986,Raup1991}. The lifetimes of species thus show a broad distribution, containing many species with short lifetimes and few species with long lifetimes. 

A very similar pattern occurs also in a spatial context: Few species inhabit large areas, whereas most species have a small range. Increasing the size of a sampling area thus leads to an increasing number of species found in that area, which is commonly known as the species-area relationship (SAR). The pattern has been studied for nearly a century now \cite{Arrhenius1921,Gleason1922,MacArthur1963,Connor1979,McGuinness1984,He1996,Tjorve2003,Drakare2006,Dengler2009}. It can be used as a powerful empirical tool, e.g. to determine the optimal sample size for an experiment or to predict the number of species in areas larger than those sampled, as suggested by Kilburn \cite{Kilburn1966}, or to set conservation targets using biodiversity survey data \cite{Desmet2004}.

Both patterns must be related \cite{Willis1922,Gaston2008}: Species with good survival chances in their local environment have more time to colonize new habitats and species inhabiting large areas have higher chances to recover after local disturbances if we assume that the spatial range of a species is the outcome of colonization and extinction dynamics. Yet, species lifetime distributions and species-area relationships have been discussed in mostly separate communities.

The exact shape of lifetime distributions is unclear because data points are few. The data are compatible with a power law, $p(t) \sim t^{-\alpha}$ (indicating self-organized criticality \cite{Drossel2001a}), as well as with an exponential decay, $p(t) \sim e^{-\nu t}$ (indicating independent and random extinction of species), or with some intermediate function \cite{Sole1996,Newman1996,Newman1999}. This ambiguity of the empirical data is also reflected in the various theoretical models for interacting species, part of which produce power-law lifetime distributions and part of which produce exponential distributions. Simple birth-death models that do not include fitness differences between species give an exponent $\alpha$ of $2$ or $1.5$, combined with an exponential cutoff \cite{Bertuzzo2011}. The simplest non-neutral models assign to species a trait value and a fitness that depends on the trait and either a global environmental variable or the traits of neighboring species. Dynamical rules include random changes in variables, extinction thresholds, and random addition of new species. Lifetime distributions of these models vary between power laws with an exponent of $\alpha=1$ \cite{Newman1997,Bak1993,Manrubia1998} and distributions with exponential tails \cite{AmaralMeyer1999,Drossel1998,Pigolotti2005}. More realistic evolutionary models, in which survival and extinction are based on the actual feeding interactions in a food-web structure, typically show only small extinction avalanches or even freeze completely \cite{Caldarelli1998,Drossel2001,LL2005,Brannstrom2011,Allhoff2015a}), unless a nonzero spontaneous extinction rate is introduced (see below). This means that lifetime distributions are broad with long mean lifetimes, and with exponential tails due to random but rare extinctions.  

The data availability of species-area relations is much better than for species lifetime distributions. A meta-analysis of 794 SAR by Drakare et al. revealed that species-area relations are significantly affected by variables characterizing the sampling scheme, the spatial scale, and the types of habitats or organisms involved \cite{Drakare2006}. The shape of empirical SAR curves is best described by a power-law or a logarithmic fit \cite{Dengler2009,Drakare2006}, but alternative curves have also been suggested, in particular to account for small island effects \cite{Lomolino2000,Triantis2003} and/or for an upper asymptote \cite{Tjorve2003}. The power-law fit, $S \sim A^z$, was first proposed by Arrhenius \cite{Arrhenius1921} in 1921, whereas the logarithmic fit, $S \sim z\cdot \ln(A)$, was proposed by Gleason \cite{Gleason1922} in 1922. In both cases, $S$ represents the number of species, $A$ represents the sampling area, and $z$ is a constant. Both authors based their model on empirical observations and both models have competed with one another for nearly a century now \cite{Drakare2006}.

The origin of species-area curves has been ascribed to random effects, such as random placement of individuals in space, or to systematic effects, such as larger areas hosting more diverse habitats or offering increased survival chances \cite{McGuinness1984}. When the spatial distribution of individuals is generated by placing individuals randomly or in some correlated manner on sites, a large variety of species-area relationships can result, depending on the used species-abundance distribution and the degree of correlation in placement\cite{He2002}. A simple patch occupancy model, where species are characterized by colonization and extinction rates, gives species-area laws with an exponent $z$ that increases with extinction rate and lies in the range of 0.5-0.8 \cite{Caswell1993}. A spatially explicit model, where individuals are removed at a constant rate and empty sites are replenished by individuals from neighboring species or from newly introduced species, gives an exponent $z$ that ranges between 0.2 and 0.8 on the regional scale. The value of this exponent is exclusively determined by the speciation rate \cite{Rosindell2007}. Qualitatively similar results were found earlier by Durrett and Levin \cite{Durrett1996}. 

All these models are so-called neutral models, which means that there is no difference in fitness between different species, and that all of them have the same chance of colonizing new sites. A non-neutral model, where each lattice site can host many individuals and where a random network of biotic interactions determines the rate at which offspring is produced, gives a similar range for the exponent, and its value  decreases with increasing migration rate \cite{Lawson2006}. When species are arranged in a food chain, the exponent $z$ depends on the placement within the food chain and increases with trophic rank \cite{Holt1999}.

For several neutral models, analytical expressions for both the species lifetime distribution and the species-area relationship were derived, establishing a relation between the values of these two exponents \cite{Maritan2016,Bertuzzo2011,Pigolotti2017}. In addition, Zaoli et al. \cite{Zaoli2017} established scaling relationships between several macro-ecological patterns, starting from size distributions of individuals in a community. However, such simple models ignore the role played by ecology and evolution: they do not explain the mechanisms behind origination and local extinction, they do not include trophic structure, and they do not take into account the fact that the interaction with other species plays an important role for species survival. 
In order to derive species-area laws and lifetime distributions from processes with a clear biological meaning, a spatially explicit, mechanistic evolutionary model is needed. The model should be (i) based on biologically interpretable species traits, (ii) complex enough to reproduce both lifetime distributions and species-area relations at different trophic levels and (iii) yet simple enough to be understandable, meaning that the different mechanisms at work can be disentangled from one another. The models mentioned so far typically combine only a subset of these features and either lack trophic structure or a trait-based evolutionary rule. 

To address these issues, we propose a multi-trophic food web model that includes both evolutionary and spatial dynamics. Predation interactions are obtained by rules that are based on a species' average adult body mass as its key trait. Evolutionary dynamics is then based on the traits that determine these predation interactions, as was done by previous authors \cite{LL2005,Brannstrom2011,Allhoff2013,Allhoff2015a}. Spatial dynamics is considered via migration over a grid of patches, each containing a local food web. This combined system of an ecological network on a spatial habitat network is a 'network on a network', or a meta-food web \cite{Gravel2011,McCann2005,Plitzko2015,Richhardt2015}.

The need to combine spatial and evolutionary dynamics has been emphasized by several authors, since gene flow and invasions potentially affect local adaptation and vice versa, depending on the relative timescale of spatial and evolutionary dynamics \cite{Kirkpatrick1997,Urban2008,Loeuille2008,Norberg2012}. While there exist already several studies of evolutionary food-web models in space \cite{Bolchoun2017,Allhoff2015,Rossberg2008}, these models do not allow the evaluation of species-area laws since their computational cost is far too high for investigating more than a handful of patches as they include explicit population dynamics. The model that we introduce has no population dynamics. Instead, species survival is determined by a numerical index that is based on the number of prey, predator and competitor species (i.e., species that have the same prey). Changes in the species composition lead to changes in the survival index, which in turn can lead to the extinction of species. In addition to these nonrandom local extinctions, our model also includes a small probability for random local extinction events, mimicking the impact of fluctuating environmental conditions. 

In some respects, our model is related to classical colonization-extinction models, which were introduced in the 1990s to model metapopulation dynamics on large scales \cite{Hanski1991,Hanski1994}, and which were in the meantime extended to larger communities \cite{Pillai2010} and also to small food web modules \cite{Pillai2011,Barter2017}. The important differences are that the majority of extinction events in our model do not happen at random but are driven by species interactions, and that we introduce new species as modifications of existing ones. With our choice of parameter values, the scales of evolution and dispersal are separated only by approximately one order of magnitude, so that changes in species composition due to evolutionary processes and those due to the influx of new species in a habitat interact with each other to generate the overall species distribution pattern. 

Our model produces a large variety of local and spatial dynamics. We investigate the resulting lifetime distributions, species-area relations, and the correlation between both. We furthermore discuss our findings both for different body mass clusters (which correlate with trophic levels) and for different model parameters reflecting environmental conditions (such as migration rate, grid size and extinction rate). By comparing our results to some of the simpler models mentioned above, we identify the features that affect the values of the power-law exponents in our model and establish a relation between them. 


\clearpage
\section{Model and Methods}
\subsection*{Overview}

We consider a grid of patches representing a spatial landscape, where each patch contains a local food web that consist of several interacting species. Each species $i$ in our model is characterized by three traits, namely its average adult body mass $m_i$, its feeding center $f_i$, and its feeding range $s_i$. These traits determine local feeding interactions: If two species $i$ and $j$ both have a population on the same patch, and if $\log{m_j}$ falls into the interval $\log{f_i} \pm s_i$, then species $i$ becomes predator of species $j$. This approach is similar to that of the well-known niche model \cite{Nische2000,Guill2008} when $\log(m_i)$ is equated with the niche value. 

\begin{figure}[ht]
\centering
 \includegraphics[width=\textwidth]{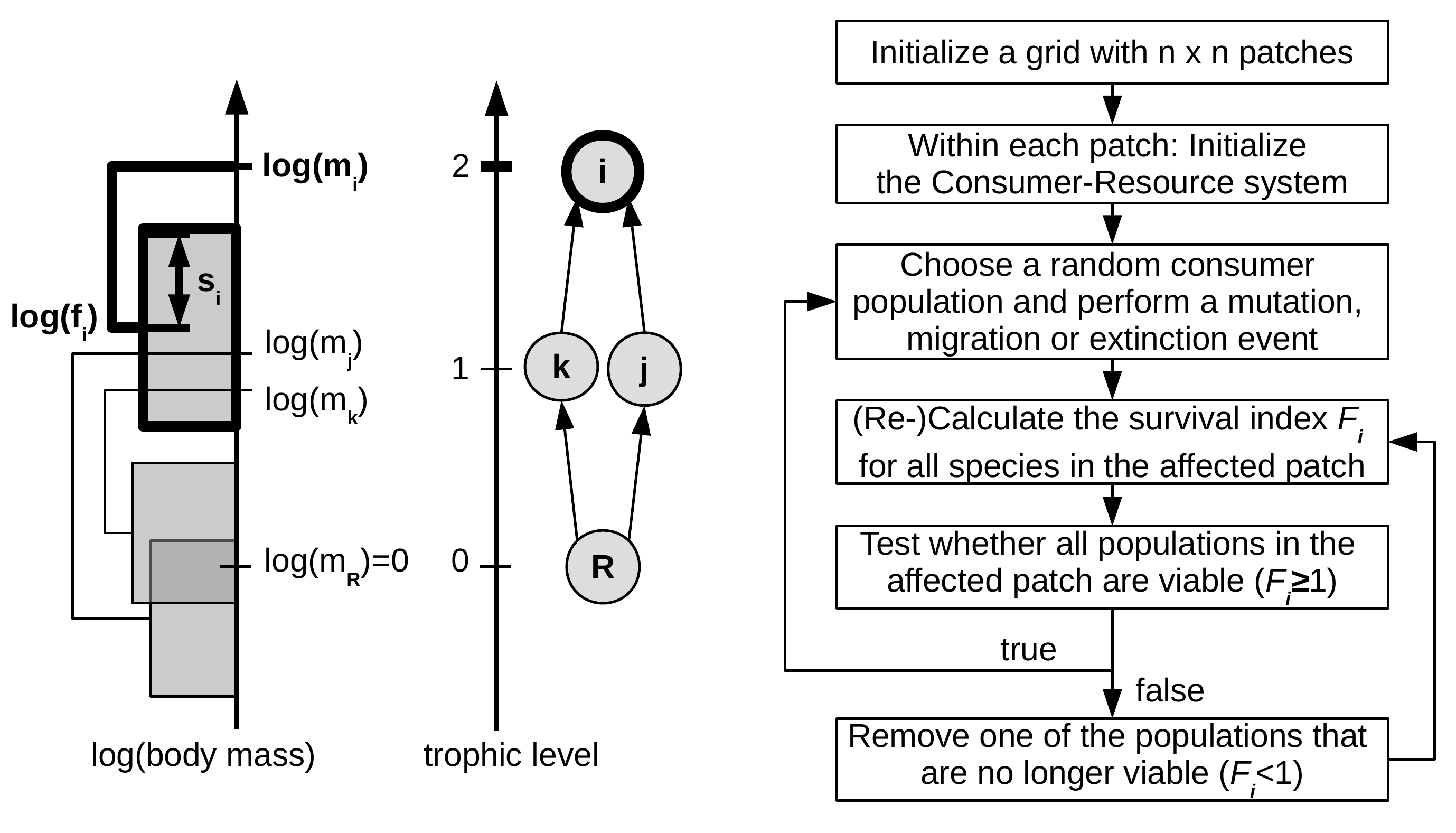}
 \caption{\textbf{Left:} Illustration of species traits and resulting feeding intervals. Species $i$ (here in bold) is characterized by its average adult body mass $m_i$ and its feeding traits $f_i$ and $s_i$. The body masses of species $k$ and $j$ fall into the interval $log(f_i)\pm s_i$, meaning that $i$ is a predator of $k$ and $j$. Energy input into the system is provided via the resource $R$, on which species $k$ and $j$ feed. Note that this food web module only serves as an illustration, as the food webs produced with this model are typically more diverse. 
\textbf{Middle:} The trophic structure of the food web module explained on the left. The trophic level $l_i$ of a species is calculated as the average trophic level of its prey plus one. 
\textbf{Right:} Overview of the simulation algorithm.}
\label{modeldescript}
\end{figure}

The local species composition within a patch is not static, but changes over time. All food webs on the grid are initialized as a consumer-resource system. New consumer species can enter a local food web either via stochastic immigration from neighboring patches, or via "mutation" events that create species that are similar to existing ones within one patch. These "mutation" events can be interpreted either as local speciation events or as changes due to genetic drift or selection forces. Species may furthermore disappear from a local food web either because they no longer have enough prey to compensate for competition, predation and mortality losses, or due to random local extinction events that mimic fluctuations in population sizes. A visualization of the species traits, as well as an overview of the simulation algorithm, is shown in Figure \ref{modeldescript}. 

\subsection*{Rules for local species survival}

The most realistic method to determine which species survive in a local food web is the calculation of population dynamics \cite{Yodzis1992}. Such coupled differential equations are typically solved using numerical methods, which can be very time consuming, especially when considering diverse communities on large spatial and temporal scales. We therefore propose a simplified approach that is based on binary networks, where all nodes (species) are either present or absent, without tracking the actual population size. We determine species survival based on a survival index $F_i$ for each species $i$. It depends on the number of its prey and predator species according to the formula 
\begin{equation}
F_i = \frac{\sum_j \alpha_{ij}}{\sum_k \alpha_{ki} + d}\, ,
\end{equation}
with a "link strength" $\alpha_{ij}$ between predator $i$ and prey $j$ that serves as a proxy for biomass flux, as explained below, and with the biomass loss term $d$. Species $i$ is considered viable if $F_i \geq 1$. 

This survival criterion is motivated by the structure of the differential equations that are typically used to describe population dynamics: 
\begin{align}
\text{growth rate} &= \sum \text{predation gain} \nonumber\\
&- \sum \text{predation loss} \nonumber\\
&- \text{mortality losses}
\end{align}
A small population (e.g. a mutant or immigrant population) is viable if the growth rate is positive and will go extinct if the growth rate is negative. Let us consider a population at the critical threshold with growth rate zero. Moving the loss terms to the left hand side of the equation and then dividing the equation by the loss terms results in the proposed form of $F_i$. 

In order to obtain an expression for $\alpha_{ij}$, we consider that biomass consumption via predation is never $100\%$ efficient, implying that a certain amount of biomass is lost at each trophic level. Links between high trophic levels are thus weaker than links between low trophic levels. We furthermore assume that a link to a given prey or resource is weaker if the predator shares it with other predators. These two assumptions are implemented in the following expression:
\begin{equation}
\alpha_{ij} = x^{l_i}\left(1-C_j\right)
 \label{eq_a_ij}, 
\end{equation}
where $x<1$ describes efficiency losses, $l_i$ is the trophic level of species $i$, and $C_j$ is a function that describes the reduction of the link strength due to interspecific competition (see below). The trophic level of a species is defined as the average trophic level of its prey plus one. Note that the trophic level of a given species depends on the local diet and might therefore differ from one patch to the next. Energy input is provided by an external resource that is modeled as a species with a fixed body mass ($m_0=1$), which cannot mutate or die. The trophic level of this resource is set to 0 by definition.

Even though this is not explicitly taken into account, we do assume that each population is not only limited via energy intake but also experiences some self-regulation via density dependent losses such as intraspecific competition. In consequence, we also assume that several predators can coexist despite sharing the same resource \cite{Chesson2000} and that the total energy uptake from some predator populations sharing a resource can be higher than the total energy uptake from a smaller number of predator populations. We furthermore assume that interspecific competition reduces the net energy gain that predators get from a shared resource, which provides an upper boundary for the maximum number of coexisting competitors. We propose that $C_j$ depends on the number of competitors $k_j$,
\begin{equation}
C_j = \frac{(ck_j)^2}{1+(ck_j)^2}  \label{eq_C_j}, 
\end{equation}
where $c$ is a model parameter that describes the overall competition strength in the system. Equation \ref{eq_C_j} describes a sigmoidal function that takes values between 0 (no competition) and 1 (maximum competition). The potential energy gain from prey species $j$ as described in equation \ref{eq_a_ij} is thus highest without competition, decreases for increasing values of c or increasing numbers of competitors, and approaches zero when the competitive pressure becomes too strong.

\subsection*{Species Turnover} 

Each simulation of the model starts with an initial configuration, where a consumer (body mass $m_1= 100$, feeding center $f_1=1$ and feeding range $s_i=0.5$) and a basal resource (body mass $m_R=1$) is present on each patch throughout the whole simulation. We consider three mechanisms leading to ongoing turnover among consumer species (see the flowchart in Figure \ref{modeldescript}). 

The first mechanism is a "mutation event", by which a new species enters the system. It occurs within one patch and can be interpreted as the origination of a new population by local evolution via adaptive changes, random genetic drift or sympatric speciation. One of the existing consumer populations $i$ is chosen at random as a "parent", which means that those patches accommodating more diverse food webs and those species inhabiting more patches have a higher probability to experience a mutation, following the idea that diversity itself favors diversification \cite{Calcagno2017}. Then, a "mutant" population $j$ is introduced into the patch of the parent. 

The logarithmic body mass of the mutant is selected at random from the interval $[\log m_i-\log q,\log m_i + \log q]$, so that parent and mutant body mass differ at most by a factor $q$. The logarithm of the mutant's feeding center, $\log f_j$, is then chosen at random from the interval $[\log m_j-f_{\text{diff,max}},\log m_j-f_{\text{diff,min}}]$, with $f_{\text{diff,min}}=0.5$ and $f_{\text{diff,max}}=3$, meaning that the typical predator-prey body mass ratio varies between approximately 3 (because $\log(3)\approx 0.5$) and 1000 (because $\log(1000)=3$). The feeding range $s_j$ of the mutant is set equal to $f_{\text{diff,min}}$. With these rules, the mutant feeds only on species with smaller body masses, such that cannibalism is excluded. A similar set of inheritance rules has been used and discussed in the evolutionary food web model by Allhoff et al. \cite{Allhoff2015a}. 

The mutant's traits determine its role in the food web and hence its ability to survive in the environment created by the other species. The mutant is considered viable if its survival index is larger than 1. Otherwise the mutation event is considered unsuccessful, and the mutant is removed from the system without further consequences. If the mutant becomes established, this can change the number of prey, predator or competitors of other species in the affected patch. We therefore recalculate the survival index of all populations within the affected patch after each mutation event. If one of the already existing species is no longer viable, meaning that its survival index $F_i$ is smaller than 1, then we remove it from the affected patch and recalculate the survival indices again. If there are several species that are no longer viable then we remove the species with the smallest value of $F_i$ (or one of those in case several species share the same survival index). The last step is repeated until only viable populations remain in the system, before the next event takes place. We would like to emphasize that it is important that only one species is removed at a time, and not all species with the smallest value of $F_i$ simultaneously, since the latter rule can wipe out an entire trophic level in one go as our model might assign to all its species the same survival index. 

The second mechanism by which the local species composition changes is a migration event, which enables species to colonize neighboring patches. Again, we choose one of the existing consumer populations as a "parent". In a next step, we introduce an "offspring" population with identical traits into one of the four neighboring patches. The offspring can establish itself if its survival index is larger than 1 (true in approximatley $20\%$ of the cases). Otherwise it is removed from the system without further consequences. If the parent species already exists in the immigration patch, the migration event also remains without consequences. A species can potentially colonize the whole grid via a series of successful migration events. Just as after a mutation event, we re-evaluate the survival indices also after each colonization event, possibly resulting in local extinction events. 

Finally, we consider rare, random local extinction events as a third mechanism. Again, one of the existing populations is chosen at random, and it is removed from the local food web. In real communities, such extinction events might occur due to intrinsic fluctuations in the population densities, which can lead to spontaneous extinction when an environmental disturbance arises. Similarly to mutation and migration events, such spontaneous extinctions can trigger further extinction events as they can change the survival indices. 

The three types of events (mutation, migration, random local extinction) take place with different rates. To define the time scale of our simulations, we set the mutation rate to 1 per patch, $\mu_{mut} = 1$. The rates of migration and spontaneous extinction per patch, $\mu_{mig}$ and $\mu_{ext}$, must then be interpreted in relation to $\mu_{mut}$. 

\subsection*{Computer simulations and evaluated quantities}

The model was simulated with a computer program written in C++, which can be made available upon request. Due to the ongoing species turnover, the system goes through many different configurations during the simulation time. This allowed us to achieve a good averaging of the quantities that we evaluated even when using only one simulation run. In order to improve statistics further, we repeated each simulation ten times. 

If not indicated otherwise, we used the parameter values given in tab. \ref{tab:mut}. The size of the system varied between 4x4 and 80x80 patches, and the simulation time was $10^6$ time units. At the beginning of the simulations, local food web diversity increases rapidly due to mutation events, and successful mutants quickly colonize neighboring patches. The system then quickly reaches a state where we observe ongoing species turnover due to an interplay of mutation, migration, and extinction events, but where the number of species stays approximately constant. We are interested in these long-term dynamics of the system. 

We evaluate lifetime distributions and species-area relationships. The lifetime of a species is simply calculated as the time difference of the first occurrence on any patch and the last occurrence on any patch. When determining the species-area relationship, a sampling method must be chosen \cite{Scheiner2003}. We sampled our patches using strictly nested areas, meaning all previously considered areas are always contained within the expanded area. Applying this to our model means that we start by determining the number of species on one arbitrary patch. For every successive data point the area under consideration is then expanded by all patches directly connected to the previously considered area and the number of species present in the considered ares is reevaluated. Due to the nature of this sampling method, the added patches per data point increase outward as concentric squares, up to a point where the diagonals of the square are equal to the width of the grid. Then, the number of added patches per data point decreases again until the species on all patches have been counted. This procedure gives the species-area distribution for a given moment in time. We averaged over 10 equally spaced time points for each simulation run, and then averaged over 10 simulations. Together, this gives an average over 100 systems. 


\begin{table}[ht]
\centering
\begin{tabular}{|c|c|c|}
	\hline
	\textbf{parameter/variable} & \textbf{letter} & \textbf{value}\\
	\hline
  	\hline
  body mass of species $i$ & $m_i$ & \\
  	\hline
  feeding center of species $i$ & $f_i$ & \\
  	\hline
  feeding range of species $i$ & $s_i$ & \\
	\hline
  survival index of species $i$ & $F_i$ & \\
  \hline
  link strength between predator $i$ and prey $j$& $\alpha_{ij}$ &  \\
  \hline
  trophic level of species $i$ & $l_i$ & \\
  \hline
  competition function for consumers of prey $j$ & $C_j$ & \\
  \hline
  efficiency parameter & $x$ & S 0.2 (0.05 - 0.3)\\
  \hline
  competition parameter & $c$ & S 0.4 (0.3 - 0.6)\\
  \hline
  mortality parameter & $d$ & S 0.001 (0.0005 - 0.005)\\
  \hline
  mutation rate & $\mu_{mut}$ & 1 \\
  \hline
  migration rate & $\mu_{mig}$ & S 10 (0 - 40)\\
  \hline
  extinction rate & $\mu_{ext}$ & S 0.01 (0.0001 - 0.1)\\
  \hline
	mutation range for body mass inheritance & $q$ & S 5 (1.5 - 6) \\
  \hline
	minimum feeding center distance = feeding range& $f_{\text{diff,min}}$ & 0.5 \\
	\hline
	maximum feeding center distance & $f_{\text{diff,max}}$ & 3 \\
	\hline
\end{tabular}

\caption{A summary of model variables and parameters. Some parameter values are fixed. For those that are not fixed, their values in the standard parameter set are given behind an S, and the intervals explored in the simulations are given in brackets. The standard set is used if not indicated otherwise. Additional results based on alternative parameter combinations can be found in the online supplementary material. 
}
\label{tab:mut}

\end{table}


\clearpage
\section{Results}

\subsection*{A typical simulation run}
\begin{figure} [ht]
\begin{tabular}{cc}
(a)
\includegraphics[width = 0.48\linewidth]{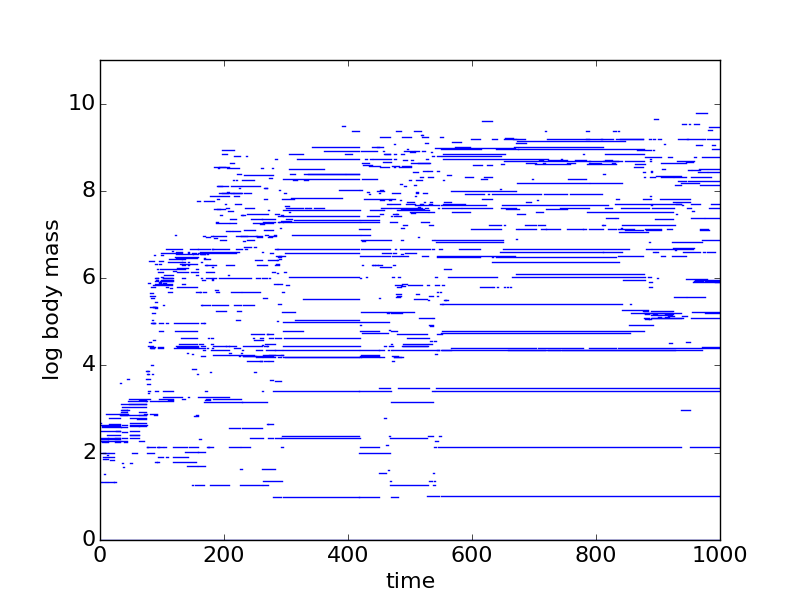}
&
(b)
\includegraphics[width = 0.33\linewidth]{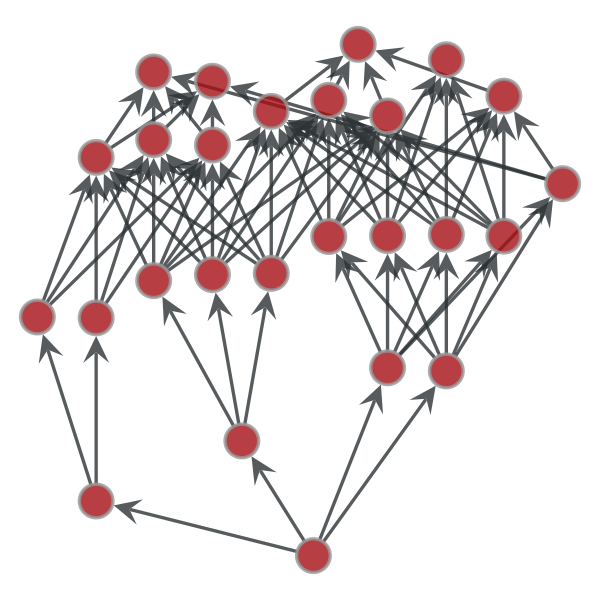}
\\
(c)
\includegraphics[width = 0.48\linewidth]{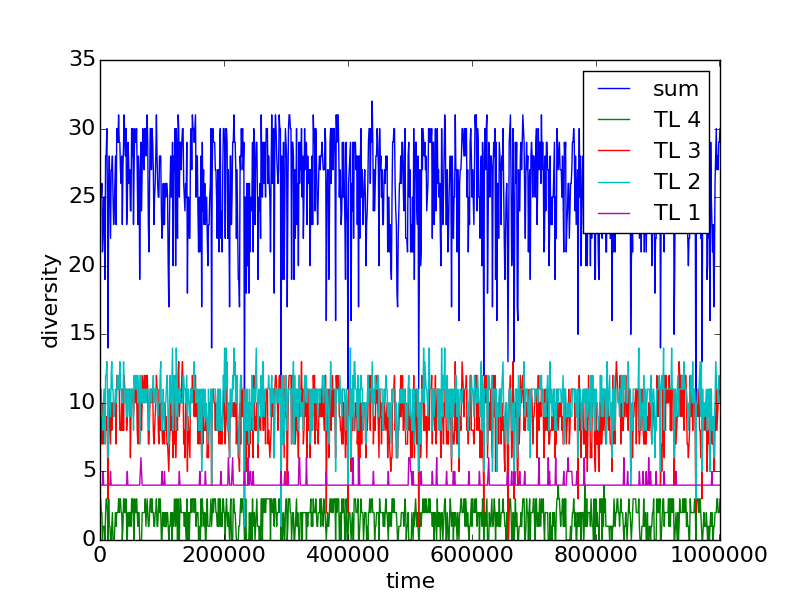}
&
(d)
\includegraphics[width = 0.48\linewidth]{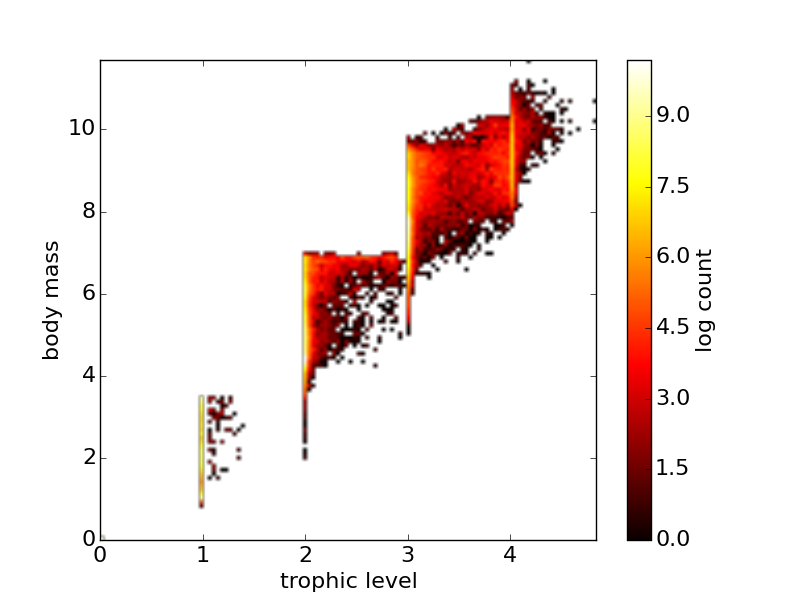}
\\
\end{tabular}
\caption{A typical simulation run using the standard parameter set:
\textbf{(a)} Time series of the body masses of all species within a given patch on a grid of 4x4 patches. Each simulation starts with a consumer-resource system within each patch, which quickly diversifies into a complex, multi-trophic food web via mutations and immigrations from neighboring patches. Note that only the initial stage is shown. Our simulations are typically $1000$ times longer. 
\textbf{(b)} The food web that emerges from the simulation in panel (a) after 1000 time units. The vertical position of a species corresponds to its body mass. Arrows point along energy flux from prey to predator.
\textbf{(c)} Number of species in one patch (in total and per trophic level) over the full simulation time. The trophic level of a species is calculated as the average trophic level of its prey plus one and can therefore vary over time and between patches. Species are assigned to distinct groups by rounding their trophic level to the next integer value. 
\textbf{(d)} The relation between body mass and trophic level of the species.}
\label{modelresults} 
\end{figure}

Figure \ref{modelresults}(a) shows the body masses of the species that are present in a given patch during the initial stage of a typical simulation run of the model. Starting with the resource and one consumer species that is present in every patch, the meta-food web is fully built up after a few hundred mutation events. There is an ongoing species turnover, with species on higher trophic levels typically living less long than those on lower trophic levels, as indicated by the length of the horizontal lines, which cover the time a species lives in the patch. The rate and size of extinction events fluctuates in time. 

A typical food web resulting from the simulation is shown in Figure \ref{modelresults}(b). It consists of four trophic layers and includes a broad range of body masses. The two intermediate trophic levels contain the largest number of species, which is also visible from Figure \ref{modelresults}(c), which shows the number of species on each trophic level over the full simulation period. This graph also shows that there are sometimes large extinction events that reduce the number of species in a patch to half its typical value. Figure \ref{modelresults}(d) shows that trophic level and body mass of a species are well correlated, with each trophic level interval being dominated by a certain body mass interval, and with most species having an integer-valued trophic level. Since the trophic level of a species depends on the other species present in the web and can therefore vary between patches, while body mass does not, we will in the following use body-mass clusters wherever the trophic-level dependence of a quantity shall be evaluated. 

\subsection*{Lifetime distributions}

\begin{figure}[ht]
\begin{tabular}{cc}
(a)
\includegraphics[width = 0.5\linewidth]{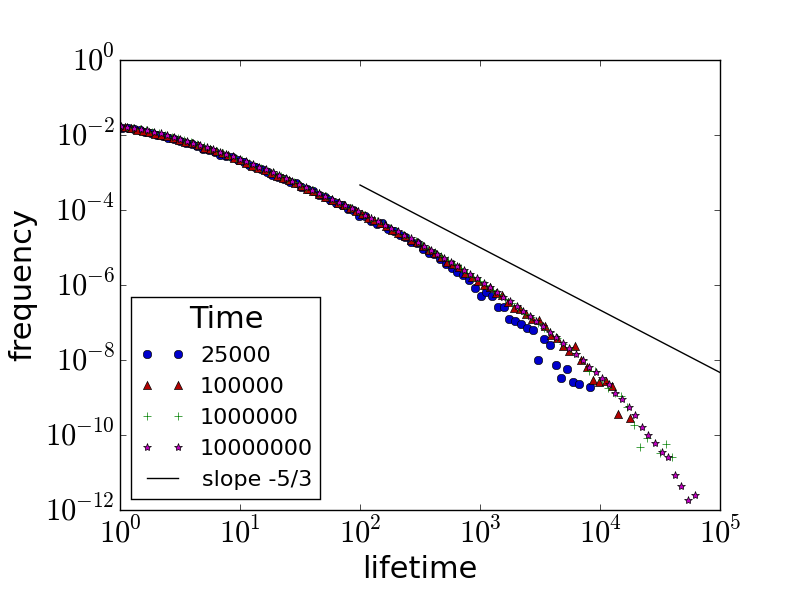}
&
(b)
\includegraphics[width = 0.5\linewidth]{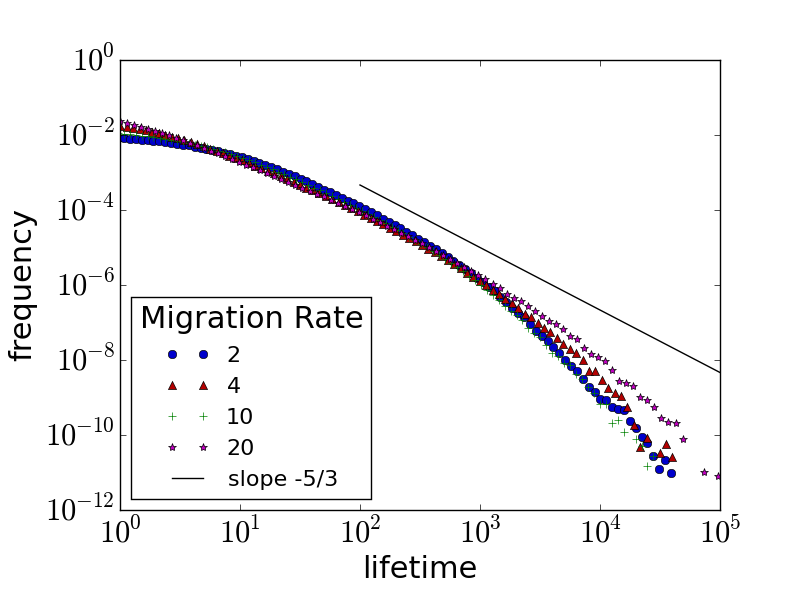}
\\
(c)
\includegraphics[width = 0.5\linewidth]{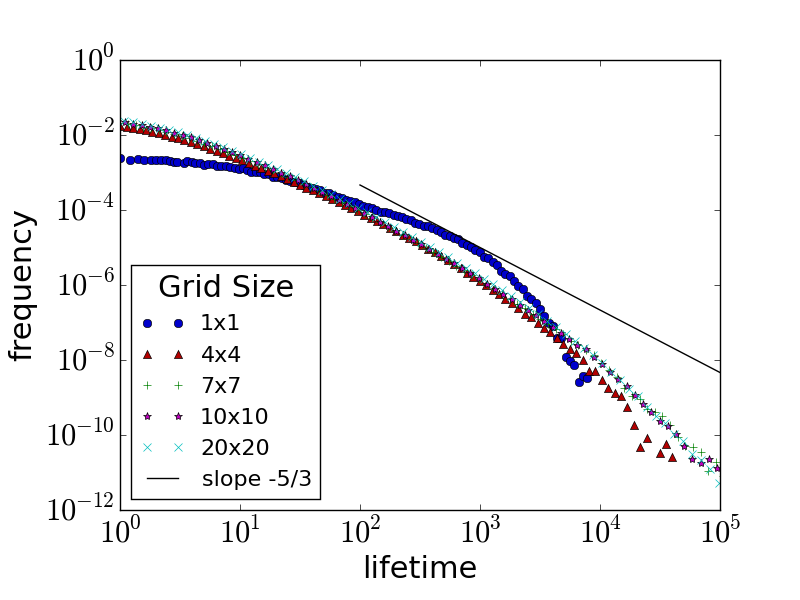} 
&
(d)
\includegraphics[width = 0.5\linewidth]{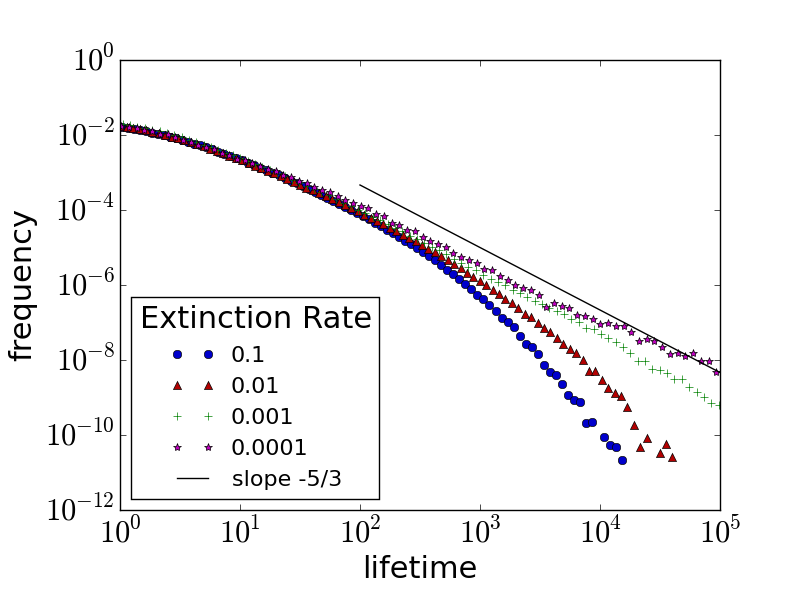} 
\\
\end{tabular}
\caption{Lifetime distribution of species for 
(a) different simulation times 
(b) different migration rates $\mu_{mig}$ 
(c) different grid sizes and 
(d) different extinction rates $\mu_{ext}$. 
Standard parameter values are used, if not indicated otherwise, as summarized in tab. \ref{tab:mut}, for a simulation time of $T = 10^6$ on a grid of size $4 \times 4$ patches. As an orientation for the eye, the straight line indicates a power law with exponent $\alpha = \frac53$. The distributions are normalized, meaning that the area under each curve has the size 1 when both axes are plotted linearly.}
\label{powerlaw2} 
\end{figure}

From the simulation run shown in Figure \ref{modelresults} it is already apparent that the lifetimes of species differ widely. In order to quantify this difference, we evaluated the lifetime distribution of all species occurring during a simulation run, see Figure \ref{powerlaw2}. This distribution is very broad and covers several decades, and for part of the parameter combinations it resembles a power law for sufficiently large times. 

The length of the simulation run mainly affects the largest lifetime seen in a plot, but not the shape of the curve, as shown in Figure \ref{modelresults}(a). This means that neither the initial built-up stage nor the final cutoff of the time series makes a significant contribution to the lifetime distributions shown in the Figure. If the initial stage contributed considerably, the shape of the curves would change with increasing runtime, and if the species that are present at the end of the simulation run were important, the curves for shorter lifetimes would show a hump towards the end of the curve. This is because species that exist at the end of the simulation time are assigned a shorter lifetime than they would have in a longer run. 

Figure \ref{powerlaw2} (b) shows the influence of the migration rate on the lifetime distribution. With increasing migration rate, the distribution becomes broader and extends to larger lifetimes. This is to be expected as species can spread further before they eventually become extinct due to the species turnover that accompanies evolutionary dynamics. This intuition will be confirmed further below in the context of species-area relationships. 

The size of the grid has only a small impact on the lifetime distribution for the parameter combinations investigated here, see Figure \ref{powerlaw2} (c). The lifetimes of the long-lived species are somewhat increased with increasing grid size, but this effect is to be expected, as the species that live longest are most likely to be present on many patches and are therefore most affected by a limited system size. There is a striking difference between the lifetime distribution on an isolated patch compared to that on a grid. This indicates that the dynamics of species turnover changes considerably in a spatial system. Indeed, on a grid migration enables by chance some species to quickly colonize many neighboring patches, which then facilitates survival even in face of local extinction events or changes in local species compositions. This effect reinforces itself, since a species that is present on several patches also has a higher chance to be selected for the next migration event. On the other hand, immigration from neighboring patches causes also many local species to lose their viability before getting the chance to spread over the grid, resulting in rather short lifetimes. In consequence, we observe both more long-lived species and more short-lived species in the spatial system, whereas intermediate lifetimes dominate in isolated systems. 

Figure \ref{powerlaw2} (d) shows the influence of the extinction rate. Lifetimes increase with decreasing extinction rate, since species persist for longer times. The lifetime distribution  follows a power law over approximately two decades with an exponent close to $-\frac53$ for an extinction rate of 0.001 for the chosen parameter set, but for most parameter sets no clear power law is visible. When we consider even smaller extinction rates, the curves bent upwards for large lifetimes, indicating that some species become so widely spread that they are likely to persist until the end of the simulation time. In order to prevent the occurrence of such "frozen" species, we chose the extinction rate in our standard parameter set accordingly. 

\begin{figure}[ht]
\centering
\includegraphics[width=0.8\textwidth]{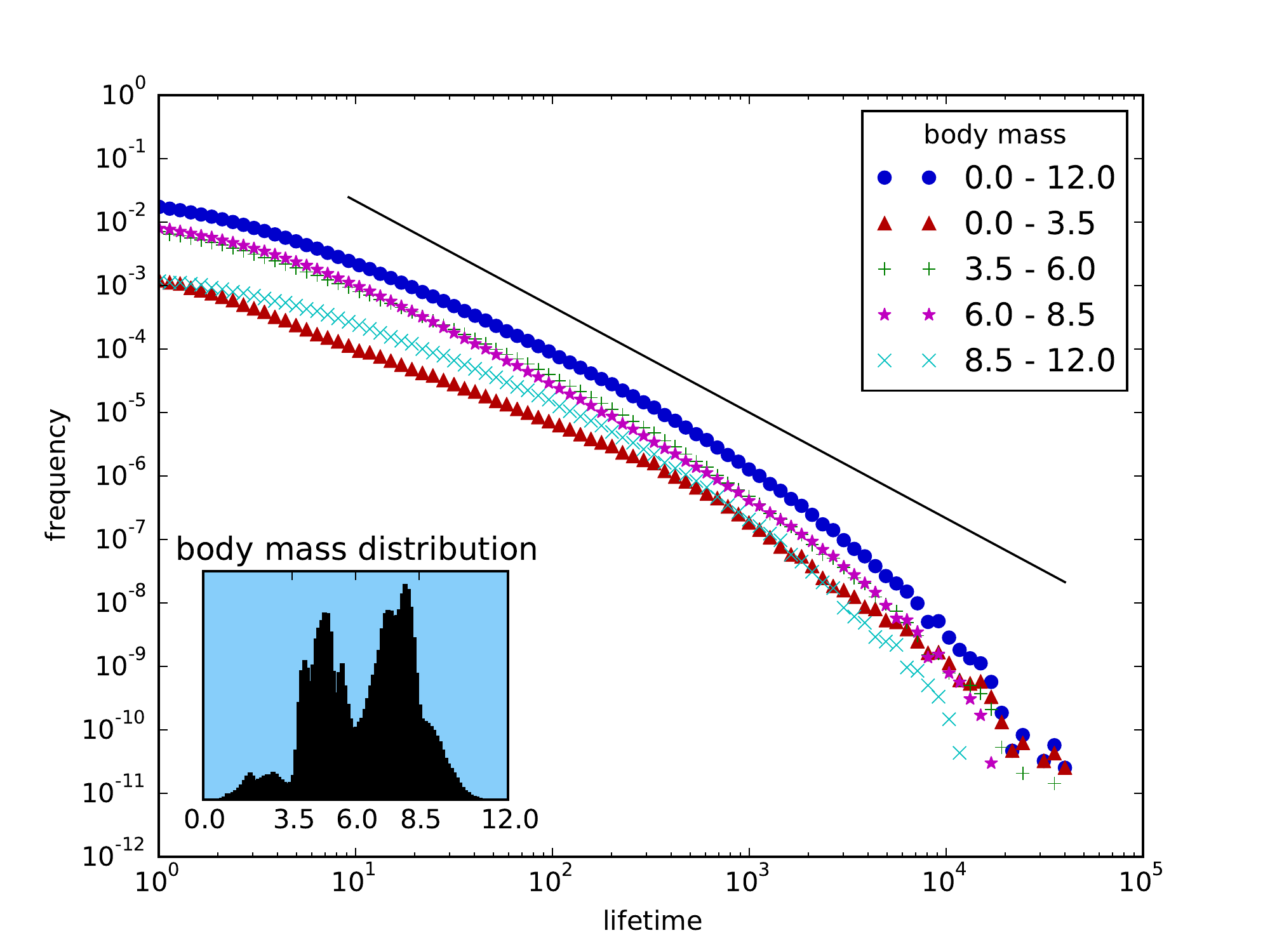} 
\caption{Lifetime distributions within the four different body-mass clusters that correspond in good approximation to four trophic level intervals. The data show that species with smaller body masses live longer on average. Simulation time is $T = 10^6$, with a grid size of 4x4. For comparison, the straight line represents a power law with an exponent of $\alpha = \frac53$. The lower curves add up to the upper curve, which corresponds to the curve representing the standard parameter set, which is also shown in all graphs of Figure \ref{powerlaw2}. Inset: distribution of body masses of all species in this simulation, showing the four body-mass clusters.}
\label{bm_dependance}
\end{figure}

Figure \ref{bm_dependance} shows how the lifetime distributions change with trophic position. In order to assign to each species a unique and integer "trophic level" value that remains constant during its entire lifetime and across patches, we made use of the correlation with (logarithmic) body mass shown in Figure \ref{modelresults}(d). This correlation enabled us to group all species that occurred during a simulation run into four body-mass clusters (see inset of Figure \ref{bm_dependance}), each of which is concentrated on one trophic level interval. The data show that the mean lifetime decreases with increasing body mass, as the curves have a larger weight at shorter times and fall off more steeply at larger times.

\begin{figure}[ht]
\centering
\begin{tabular}{cc}
(a)
\includegraphics[width = 0.5\linewidth]{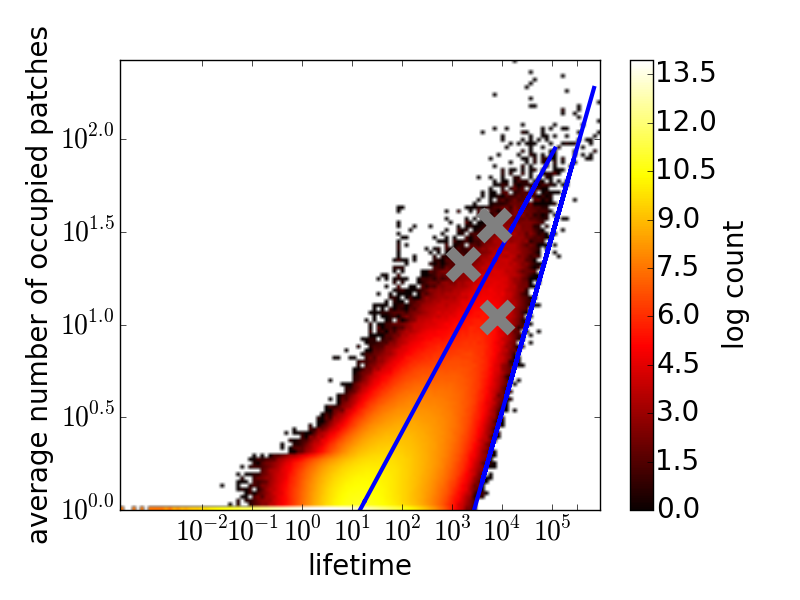}
&
(b)
\includegraphics[width = 0.5\linewidth]{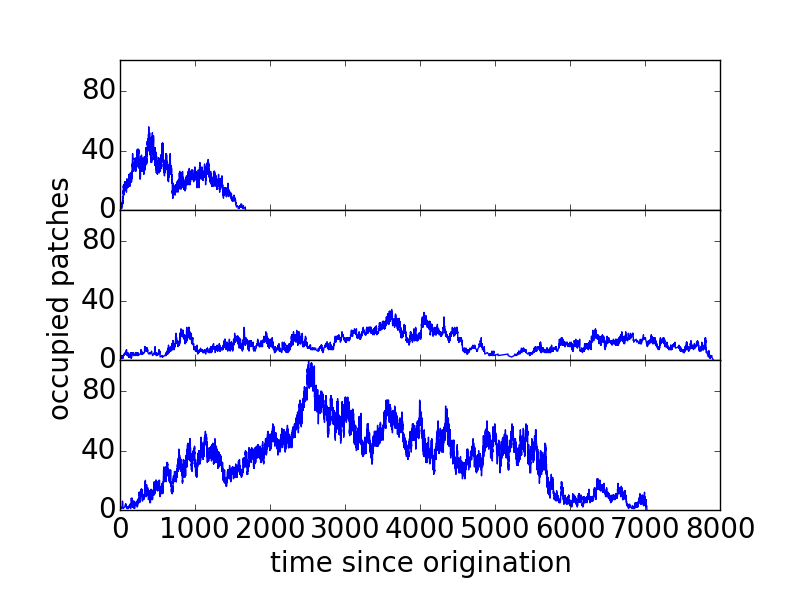} 
\end{tabular}
\caption{Simulation with standard parameters on a 20 x 20 grid. (a) Relationship between lifetime and average number of occupied patches for all species that occurred during the simulation run. The straight lines show power law fits through mean and maximum lifetimes. Their slopes are 0.499 and 0.949 with fit errors $ < 10^{-3}$. 
(b) The dispersal of three example species (indicated by crosses in (a)) during their lifetime.
}
\label{dispersal1}
\end{figure}

We explored the connection between the lifetime of a species and the average number of patches it occupies. Figure \ref{dispersal1}(a) shows the results for a typical simulation run using the standard parameter set. While the average number of occupied patches can vary widely for a given lifetime, there is of course a positive correlation between the lifetime and the number of occupied patches. 

The two straight lines are power-law fits to the mean lifetimes (when for each y value the x values are averaged) and the maximum lifetimes (when for each y value the maximum of x is chosen). Their slopes are 0.499 and 0.949. These values are close to the values 0.5 and 1, which we expect based on simple arguments: On average, each species is equally likely to lose a patch or to conquer a new one during a time step, which means that the number of occupied patches behaves like a random walk, leading to an exponent 1/2. For a given average number of occupied patches the lifetime of a species is minimized if it grows constantly until reaching its maximal spread, and then retracts again. This leads to an exponent of 1 for longest living species. 

Figure \ref{dispersal1}(b) shows the time series of the number of occupied patches for three selected species. These species spread during a relatively short time over many patches as they move into their niche in a group of connected patches. Then their range keeps growing and shrinking as the evolutionary process in the system continues and affects their local survival chances in unpredictable ways. The three corresponding lifetimes and average number of occupied patches are indicated with crosses in Figure \ref{dispersal1}(a).

\subsection*{Species-area relationships}

\begin{figure}[ht]
\centering
\includegraphics[width = 0.8\linewidth]{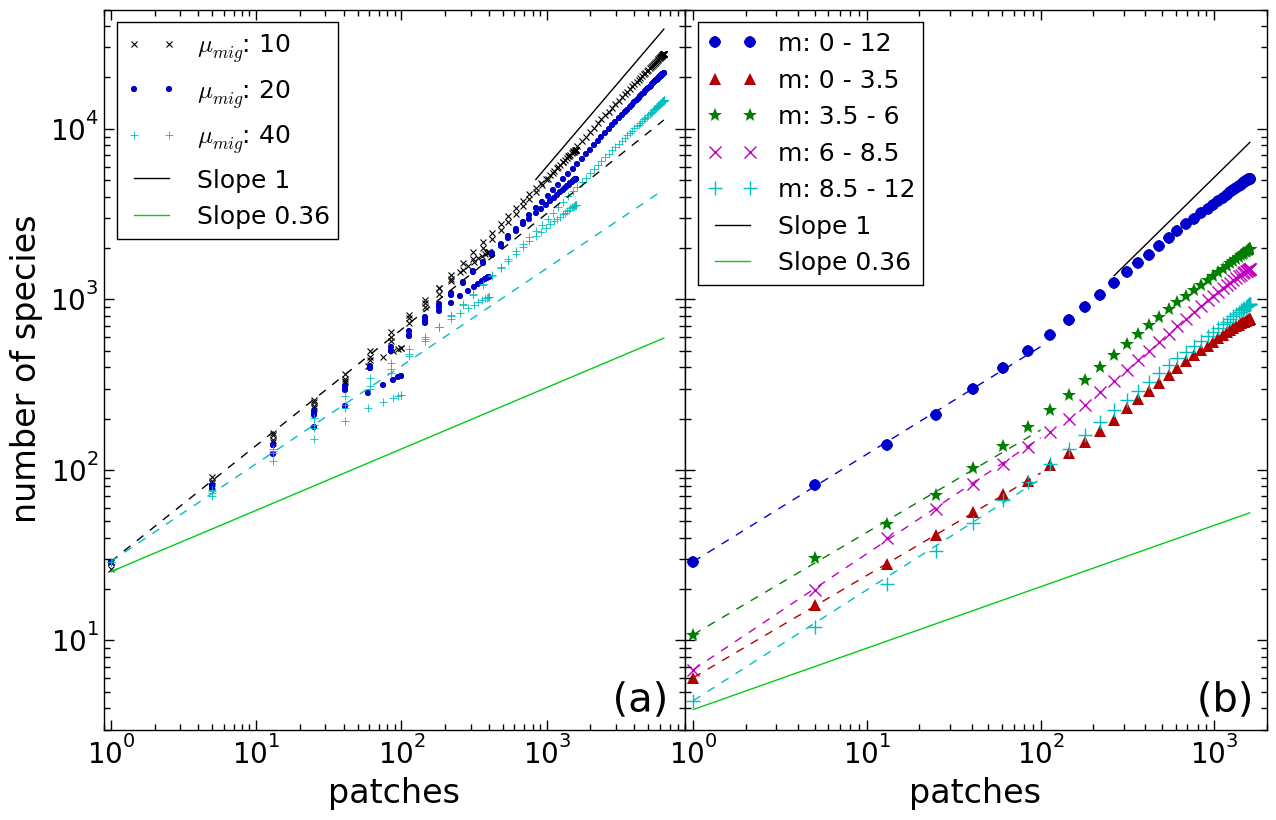} 
\caption{(a) The SAR for varying grid sizes and migration rates. The dashed lines are power-law fits over the first part of the curves (up to 100 patches), giving the exponents 0.65, 0.63, 0.59 for migration rates 10, 20, 40. (b) SAR for different body mass clusters (corresponding to trophic level intervals) with migration rate $\mu_{mig} = 20$. Here, the the exponents are 0.63 for the top curve and 0.60, 0.60, 0.68, 0.65 for clusters with increasing body mass. The fit errors in each case are less than $5\cdot 10^{-4}$.
A slope of 1 and 0.36 are given as references, where 1 is the expectation value for large grid sizes and 0.36 an empirical reference value \cite{Drakare2006}. }
\label{Fig_SAR_Plots_komb}
\end{figure}

Figure \ref{Fig_SAR_Plots_komb} shows the Species-Area Relationship (SAR) for different migration rates and grid sizes. The migration rates can be distinguished by the different markers and colors, whereas the curves for different grid sizes are identifiable by the x values of their end points and the increased density of the points when approaching the full grid size. As the system size increases, the curve for each migration rate approaches an asymptotic shape that is characterized by an increasing slope. Due to finite-size effects, the curves for smaller grid sizes branch off from this asymptotic curves when the number of occupied patches approaches its maximum possible value. When analyzing the species-area law of the curves this portion is disregarded. For patch numbers up to about 100, the curves are approximately linear on the given log-log scale, in accordance with a power law. The slope of the curves in this region with a grid size of 80$\times$80 patches were determined for migration rates of 10, 20 and 40 to be 0.65, 0.63 and 0.59 respectively. With increasing migration rate, the slope in this region decreases. Since the average number of patches inhabited by an individual species increases with the migration rate, but the number of species supported by a single food web remains the same, diversity decreases and leads to a less steep SAR. 

Beyond this power-law regime, the curves bend upwards, and the slope of the curve approaches unity. This is the expected behavior since any given species can only have a finite area of influence on the grid. When the number of considered patches becomes very large, each species occupies only a small proportion of patches,  and the number of species increases linearly with the number of patches. In addition to the two regimes (regional scale with a nontrivial power law, continental scale with a slope of 1), species area-curves usually also have a steeper slope for small areas, i.e., on the local scale \cite{Rosindell2007}. However, since our model is not individual-based and since patches represent an ecosystem on the regional scale, we do not resolve the local scale.

The general features of the species-area relationships mentioned so far can also be seen when the data are split into different body-mass clusters (which again correspond to trophic level intervals), as shown in Figure \ref{Fig_SAR_Plots_komb}(b), but with quantitative differences. While the curves for body-mass clusters in the middle of the body mass range are similar to the curve considering all species, the body-mass clusters for small and large species show a significant difference in slope, thus leading to an intersection of the two curves. Due to the food web structure, species with higher body mass are more susceptible to (secondary) extinction than species in lower trophic levels, leading to more fluctuation and thus higher diversity and a steeper curve.

\begin{figure}[ht]
\centering
\includegraphics[width = 0.8\linewidth]{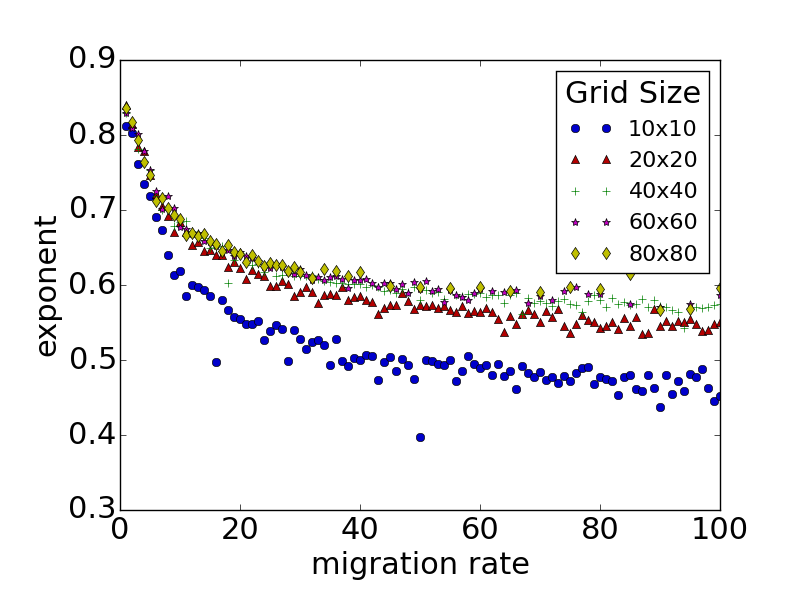}
\caption{Dependency of the exponent $z$ of the SAR on the migration rate for different grid sizes. The exponent decreases with migration rate and increases with grid size. Error bars due to fitting are smaller than the shown markers. Due to the long simulation times required for larger grid sizes and larger migration rates, the figure contains less data points for the two largest grid sizes.
\label{fig_SAR_z}}
\end{figure}

In order to obtain an overall view of the effect of grid size and migration rate on the SAR in our model, the slope in the region with smaller patch numbers was determined for different grid sizes and migration rates, see Figure \ref{fig_SAR_z}. A steep decrease in slope of the species-area curves can be observed for an increase of the migration rate when migration rate is low. For higher migration rates, the effect of increasing the migration rate on the slope becomes less relevant. This agrees with the conclusion in \cite{Rosindell2007} that the dependence of the slope on migration rate vanishes for large migration rates. The dependence on grid size is due to finite-size effects, which vanish when grid sizes are sufficiently large. 

\subsection*{Connection between lifetimes and species-area relationships}
In our model we find that the lifetime distributions take the form of a power law with an exponent close to $-5/3$ for certain parameter ranges. Simultaneously, species-area relations take the form of nontrivial power laws with exponents around 0.6 for sufficiently large migration rates and grid sizes. 

Both patterns are connected via the dynamics of our model and arise simultaneously. We find that those species that occupy on average $a$ patches (when averaging over their lifetime), have a mean lifetime that increases close to quadratically with $a$ (see Figure \ref{dispersal1}(a)). We can in fact relate this quadratic law to the above two power laws by employing scaling arguments similar to those suggested by other authors \cite{Bertuzzo2011,Maritan2016,Pigolotti2017,Zaoli2017}:
If the lifetime distribution is given as 
$p(t) \sim t^{-\alpha}$
and the relation between lifetime and area as 
\begin{equation}
a\sim t^b\, ,
\end{equation}
then the lifetime distribution of those species that exist at the same time scales as 
\begin{equation}
p_0(t) \sim t^{1-\alpha}\, .
\end{equation}
The distribution of their areas $p_0(a)$ is obtained from $p_0(t)dt=p_0(a)da$, giving
\begin{equation}
p_0(a) \sim a^{\frac{2-\alpha}b-1}\, .
\end{equation}
Now let $S(A)$ be the average number of species in an area $A$. Only species with $a<A$ can be fully contained in this area. Each of these species has the same probability to be found in this area. The number of species contained in area $A$ increases therefore with $A$ in the same way as the number of species with $a<A$, leading to 
\begin{equation}
S(A) \sim \int_0^A p_0(a)da \sim A^{(2-\alpha)/b} \equiv A^z\, .\label{scalingrelation}
\end{equation}
Using $2-\alpha \simeq 0.3$ and $1/b \simeq 2$, we obtain $z \simeq 0.6$, in agreement with our simulation results. (As this calculation does not take into account cutoffs to the power laws, its results are valid only as long as scaling regions are large enough.)


\clearpage
\section{Discussion}
We introduced a mechanistic food web model that contains both evolutionary and spatial dynamics. Despite all simplifying assumptions, in particular concerning the proposed species survival index $F_i$, the model produces complex, multi-trophic food web structures with ongoing species turnover. It provides a powerful framework to investigate how the underlying spatio-temporal mechanisms shape the observed lifetime distributions, species-area relations, and the correlation between both. In contrast to models that perform explicit population dynamics in order to determine which species die out, our model requires much less computer time and can therefore be used to study food webs on long temporal and large spatial scales. We find that the lifetime distributions take the form of a power law $p(t) \sim t^{-\alpha}$ with an exponent close to $\alpha=\frac53$ for certain parameter ranges. Species-area relations take the form of nontrivial power laws with exponents slightly below $z=0.6$ for our largest migration rates and grid sizes. 

We furthermore find that the exact shapes of the lifetime distributions and the species-are relations depend on the position in the food web. Species with large body masses and high trophic positions have shorter lifetimes than species with small body masses and low trophic positions. This can be explained with an increasing risk of secondary extinctions along the food chain, that is, with trophic position. Species with long lifetimes can colonize more patches. In consequence, we observe flatter (steeper) species-area curves for species with smaller (larger) body masses, in consistency with the prediction derived from a simple community model composed of “stacked specialist” food chains \cite{Holt1999} and with empirical observations \cite{Drakare2006}.

Our model results are in line with previous findings from empirical and theoretical studies concerning the lifetime exponent $\alpha$. The comparison is, however, not straight forward, because of the wide range of values reported in the literature. If the reported lifetime distribution follows a power law at all, simple mathematical models give values between $\alpha=1$ and $\alpha=2$, as reported in the Introduction, and empirical data lie in the same range, with, e.g., $\alpha \sim 1.6$ \cite{Keitt1998} (which, however, is obtained from local lifetime distributions and on short timescales) or $\alpha=1.8$ \cite{Bertuzzo2011} for breeding birds. These values are close to the one appearing in our simulations. The fact that $\alpha$ lies between 1 and 2 implies that the average lifetime of species, which is given by the integral $\int_0^{t_{max}} t p(t)dt$ depends on the upper cutoff of the integral. This cutoff can be determined by major extinction events or other rare or slow processes that cause species replacement but do not affect directly the exponent $\alpha$. Slow processes are often included in $p(t)$ not as a sharp cutoff but as a soft cutoff by writing $p(t) \sim t^{-\alpha}e^{-t/t_{max}}$. In our model, this cutoff results from a combination of finite simulation time, nonzero spontaneous extinction rate, and finite grid size. 

Similarly, the values for the SAR exponent $z$ obtained from empirical data or mathematical models vary widely. Drakare et al. (2006) report that the average value of $z$ for empirical power-law relationships across various data sets is $0.27$ \cite{Drakare2006}. Considering the impact of different census methods and taking only nested sampling into account, as we did in our study, results in a steeper average slope of 0.36. Both values are, however, much smaller than the slopes that result from our model, as shown in Figure \ref{fig_SAR_z}. 

As mentioned in the introduction, simple mathematical models can generate a wide range of values for $z$ depending on the model parameters. In particular, it is possible to decrease the exponent to any small positive value in a neutral model \cite{Pigolotti2017}. For this reason, we consider it possible that our model would also yield exponents of the order of the empirical values if the migration rate was increased considerably. However, this would require much larger grid sizes (to avoid finite-size effects) and much longer simulation times than those accessible to us. Interestingly, the tangled bank model, which is not a neutral model, gives an exponent $z$ very close to the above empirical value $0.27$ \cite{Lawson2006}. However, the authors do not discuss how this exponent varies when model parameters are changed. 

In any case, the relation (\ref{scalingrelation}) between the different exponents  that we derived further above reveals that the value of $z$ depends sensitively on $2-\alpha$ and $b$, which characterize lifetime distributions and the relation between average range and lifetime. With an exponent $\alpha=1.82$ instead of 5/3, for instance, our $z$ value would move down to 0.36. The same would happen with $b=0.91$ instead of 1/2. The value of $b$ of our simulations is essentially determined by the way how species spread and become replaced on the grid. In our model, patches are added or removed from the range of a species with similar probabilities throughout the lifetime of the species. In contrast, in nature, species often show an initial period of expansion (during which the probability of addition of a patch to the species range is increased), followed by a later period of retraction (during which the probability for removing a patch from the species range is increased) \cite{vzliobaite2017reconciling}. Such a process would increase $b$ to a value closer to 1. With the stable environmental conditions presupposed in our model (no temporal parameter variation, no spatial heterogeneity), such processes are not captured. 

Also not taken into account, but possibly relevant in this context, is the potential overlap of the ecological and the evolutionary time scale \cite{hairston2005rapid,Pantel2015}. In empirical systems, population dynamics changes the proportions of the different genotypes in the system and thus couples the ecological and the evolutionary time scales. Since our model does not include population dynamics, it cannot capture this effect. For example, a threatened population might in reality start to adapt to changing conditions and this adaptation might, if it happens fast enough, ensure survival. Such "evolutionary rescue" \cite{Gomulkiewicz1995} is impossible in our model, since we instantaneously remove populations with survival indices below 1 from the system before performing the next migration, mutation or extinction event. We assume that taking evolutionary rescue into account would lead to longer lifetimes and hence flatter SAR curves. 

An important feature that is present in our model but absent in most other models is the trophic structure of the networks. Most existing models are neutral or based on a community of species that are on the same trophic level. Even the more complex tangled-bank model \cite{Lawson2006} does not contain a trophic ordering. By contrast, the species turnover rates in our model are not put in by hand but emerge more realistically from the combined ecological and evolutionary rules and consequently differ on different trophic levels. We found that even when taking into account that local processes of speciation and extinction are context-dependent instead of random, we get similar distributions as with the simple neutral models. To a certain extent, our results thus validate earlier modeling approaches. But our model also leads to new insights that can not be gained from neutral models. For example, the relation between lifetimes and number of occupied patches (shown in Fig. 5) varies widely, reflecting contextual effects in our model that vary in space, time, and with trophic position. We find that species on the second trophic level are the most numerous and therefore dominate the species-area exponent. It is known that this exponent increases with trophic level \cite{Holt1999}, and it is therefore natural that our model gives a larger exponent than models or empirical studies that consider only one low-lying trophic level. 

Our model is certainly oversimplified and it is difficult to underpin those abstract model parameters with empirical values. We therefore performed extensive robustness checks to understand whether and how changes in the model details affect the emerging lifetime distributions and species-area curves, which are available in the online supplementary material. To this purpose, we varied the ecological parameters (by varying the efficiency $x$, the competition strength $c$ and the loss rate $d$) as well as the evolutionary rules (by varying the range and shape of the mutation interval, i.e., the similarity between parent and mutant body mass). 

We find that the ecological parameters affect model results in mostly predictable ways: an increase in the efficiency $x$, a decrease in the competition strength $c$ or a decrease in the loss rate $d$ facilitates species survival and hence leads to broader lifetime distributions. These patterns are valid over a considerable parameter range as long as the emergence and persistence of multiple trophic levels is still possible. Deviations from these trends occurring for networks with few trophic levels (e.g. because of too low efficiency, too strong competition or too high loss rates) are explained in the supplementary material. 

We furthermore find that the mutation range, which is encoded by the factor $q$ describing the maximum body mass ratio between a parent and its mutant species, has only a minor impact on the resulting food web structures, species lifetime distributions and species-area relationships, if $q$ is sufficiently large. Only when the mutation range is small compared to our standard parameter, is the build-up of multiple trophic levels suppressed. Apart from that, we find that our model is also robust against changes in the shape of the distribution (e.g. Gaussian instead of uniform) from which the mutant body mass is sampled.


\begin{acknowledgements}
The bachelor thesis of Johannes Reinhard contributed to the initial stage of this study by demonstrating that including a spontaneous extinction rate is essential for obtaining an ongoing species turnover. 
\end{acknowledgements}


\clearpage
\appendix
\bigskip
\setcounter{page}{1} 
\renewcommand\thefigure{\Alph{section}\arabic{figure}}
\setcounter{figure}{0}
\setcounter{section}{19}
\section*{Supplementary Material}

\subsection*{Variation of the ecological parameters}

In the following, we discuss how the efficiency $x$, the competition strength $c$ and the mortality losses $d$ affect the network build-up, lifetime distributions and species-area relations in our model. All other parameters are chosen as in tab.~\ref{tab:mut}. 

\begin{figure}[h]
\begin{tabular}{cc}
(a)
\includegraphics[width = 0.45\linewidth]{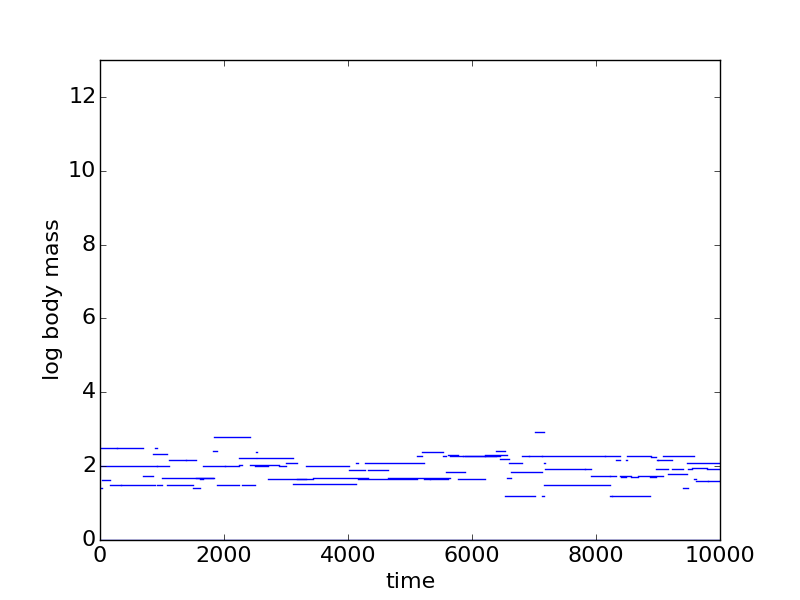}
&
(b)
\includegraphics[width = 0.45\linewidth]{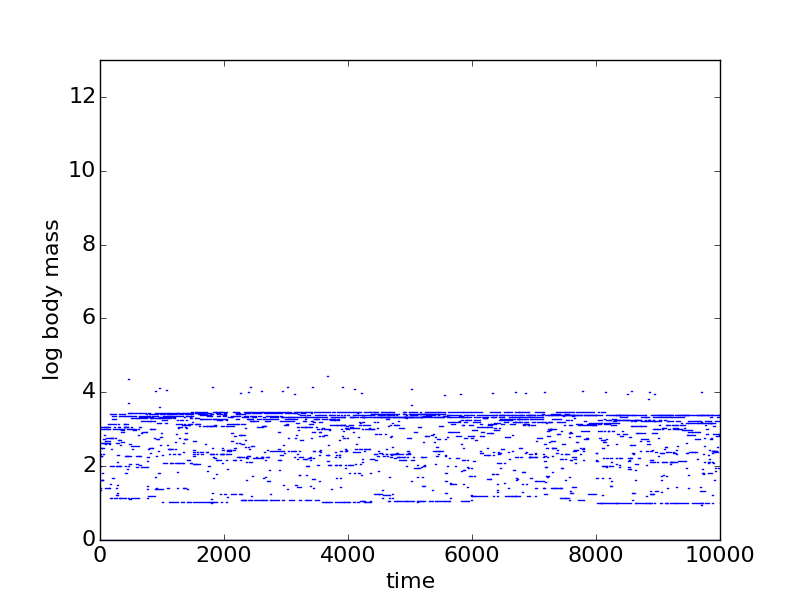}
\\
(c)
\includegraphics[width = 0.45\linewidth]{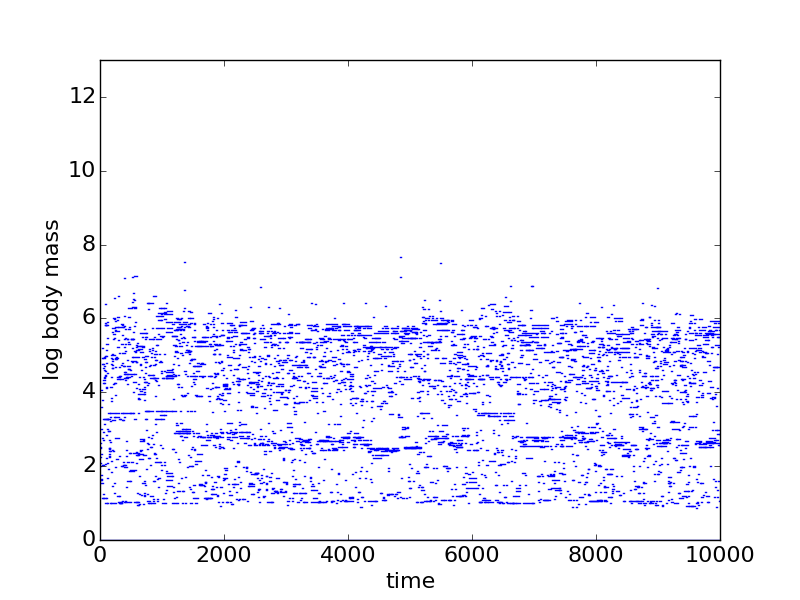}
&
(d)
\includegraphics[width = 0.45\linewidth]{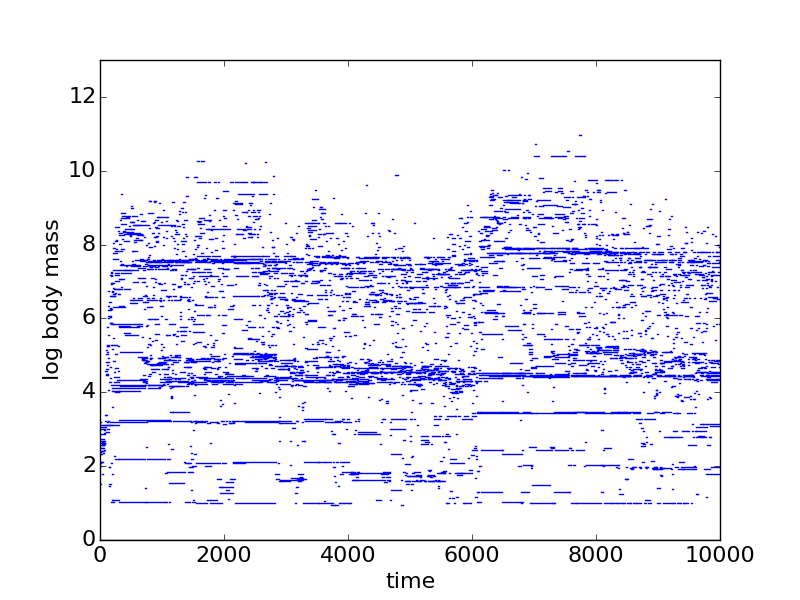}
\\
(e)
\includegraphics[width = 0.45\linewidth]{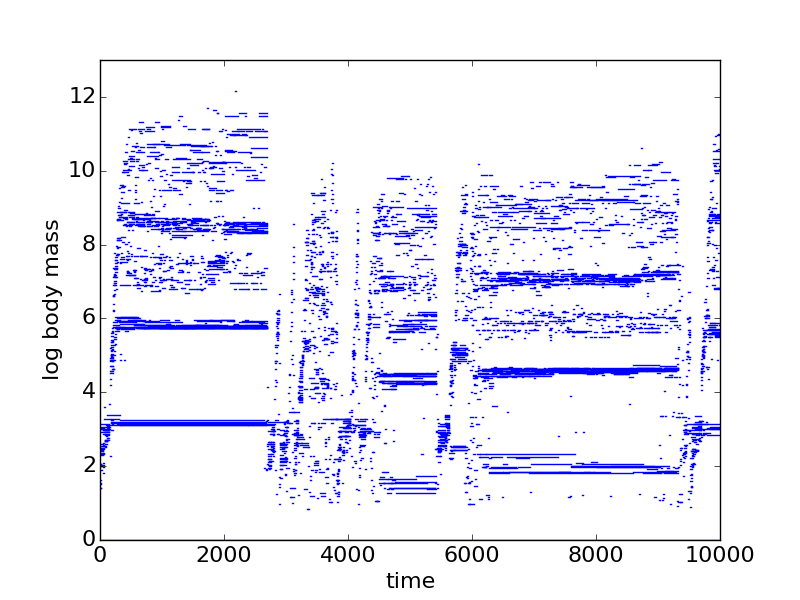}
&
(f)
\includegraphics[width = 0.45\linewidth]{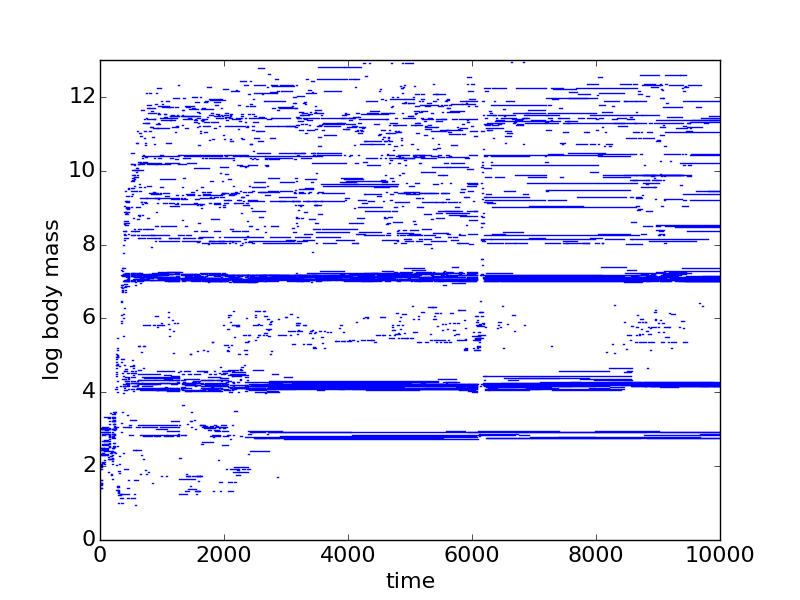}

\end{tabular}
\caption{Evolution of body masses within exemplary simulations runs using different values for the efficiency parameter $x$: (a) $x=0.05$ (b) $x=0.1$ (c) $x=0.15$ (d) $x=0.2$ (e) $x=0.25$ (f) $x=0.3$. A line indicates the presence of a species from its first emergence on this patch until its local extinction, whereas dots represent species that emerged (via mutation or immigration) but were not able to survive.}
\label{x_timeseries}
\end{figure}

\begin{figure}[p]
\begin{tabular}{cc}
(a)
\includegraphics[width = 0.45\linewidth]{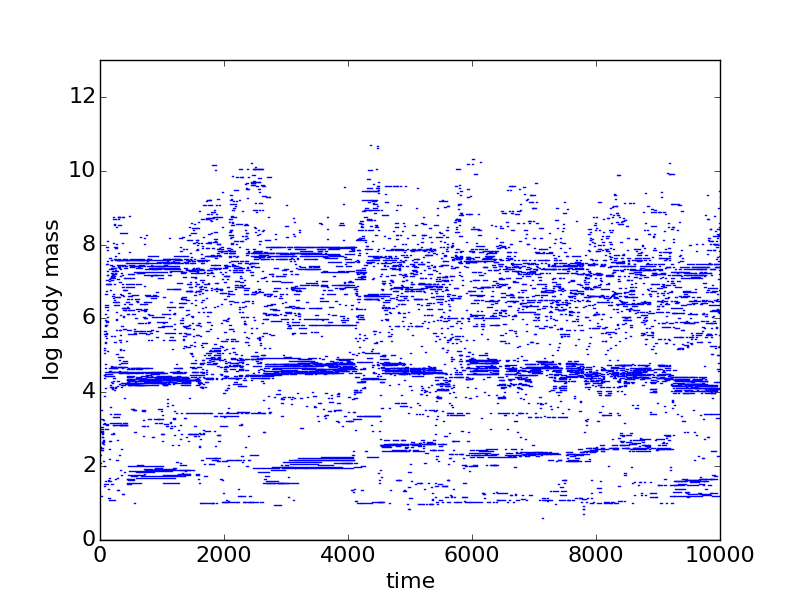}
&
(b)
\includegraphics[width = 0.45\linewidth]{Attachment/timeseries_S-1M_m4_4x4_-2_p0_T0_0-10000_0.png}
\\
(c)
\includegraphics[width = 0.45\linewidth]{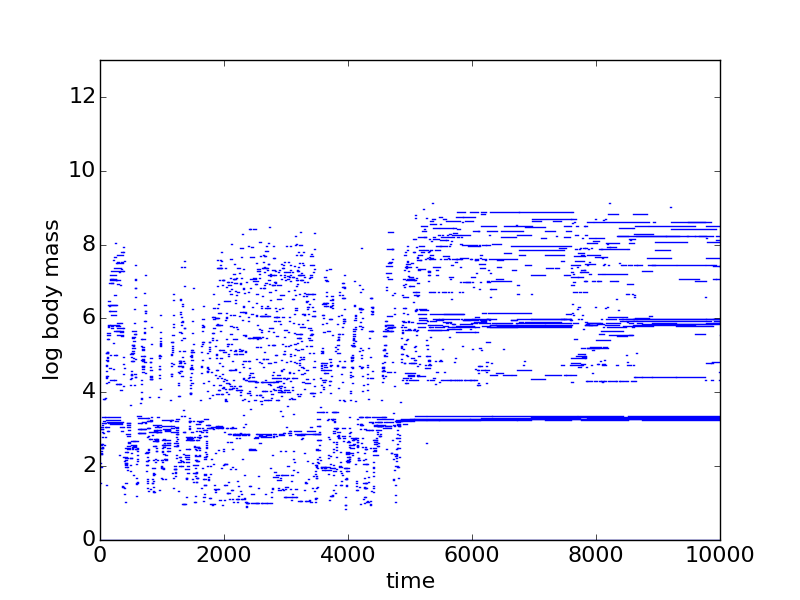}
&
(d)
\includegraphics[width = 0.45\linewidth]{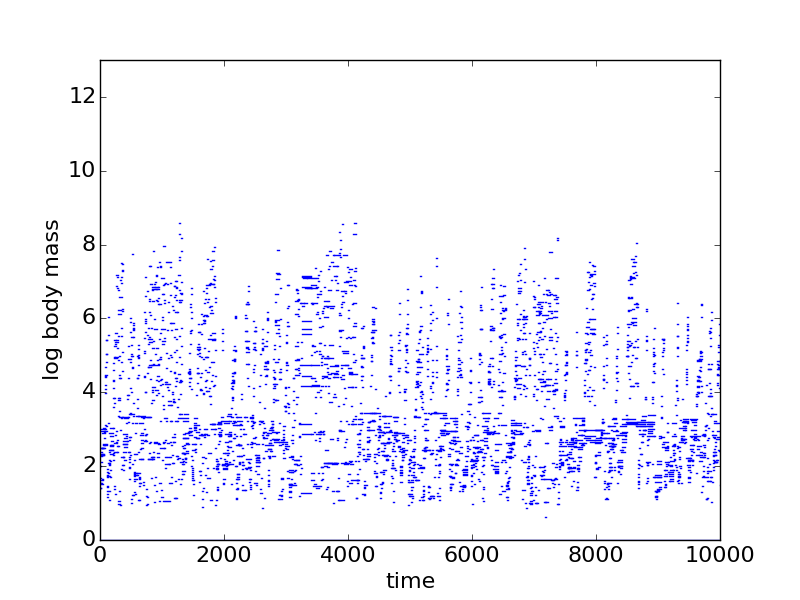}
\\
\end{tabular}
\caption{Evolution of body masses within exemplary simulations runs using different values for the competition parameter $c$: (a) $c=0.3$ (b) $c=0.4$ (c) $c=0.5$ (d) $c=0.6$}
\label{c_timeseries}
\end{figure}

\begin{figure}[p]
\begin{tabular}{cc}
(a)
\includegraphics[width = 0.45\linewidth]{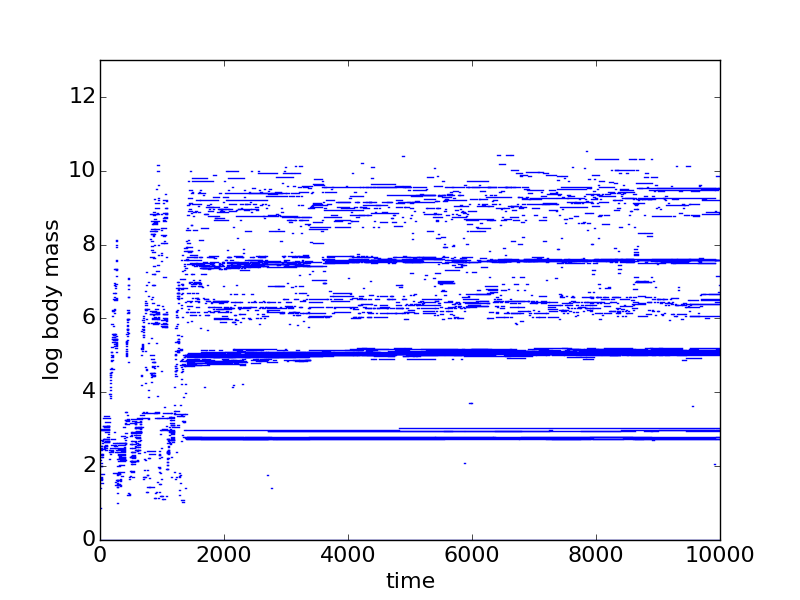}
&
(b)
\includegraphics[width = 0.45\linewidth]{Attachment/timeseries_S-1M_m4_4x4_-2_p0_T0_0-10000_0.png}
\\
(c)
\includegraphics[width = 0.45\linewidth]{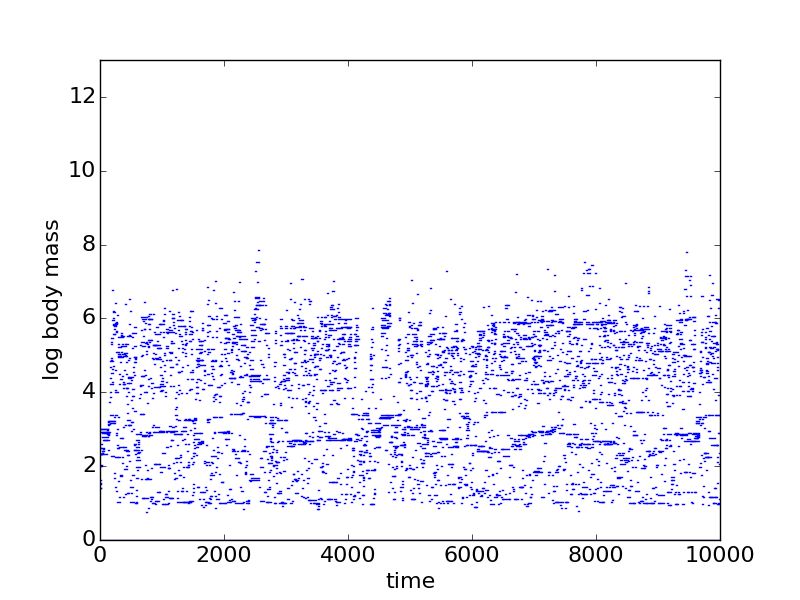}
&
(d)
\includegraphics[width = 0.45\linewidth]{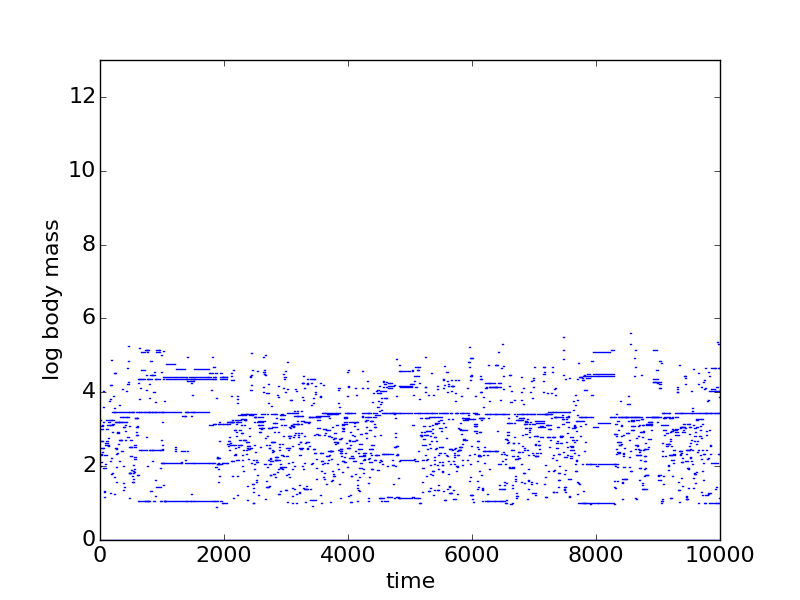}
\\
\end{tabular}
\caption{Evolution of body masses within exemplary simulations runs using different values for the mortality parameter $d$: (a) $d=5\cdot 10^{-4}$ (b) $d=10^{-3}$ (c) $d=2\cdot 10^{-3}$ (d) $d=5\cdot 10^{-3}$}
\label{d_timeseries}
\end{figure}

We find that $x$, $c$ and $d$ affect species co-existence and hence the emerging food web structures in very similar ways: Higher efficiencies, weaker competition and lower loss rates facilitate species survival, which in turn enables more species to coexists within a given trophic level, which then increases the survival chances in the next trophic level. In consequence, we observe more diverse systems and bigger body masses (which is correlated to high trophic positions, see Figure \ref{modelresults}) in systems with high efficiency (Figure \ref{x_timeseries}), weak competition (Figure \ref{c_timeseries}) and low loss rates (Figure \ref{d_timeseries}). Consequently, we also observe that the emergence of multiple trophic levels is hampered in cases where the efficiency is too low, the competition is too strong or the loss rates are too high. 

\begin{figure}[p]
\begin{tabular}{cc}
(a)
\includegraphics[width = 0.45\linewidth]{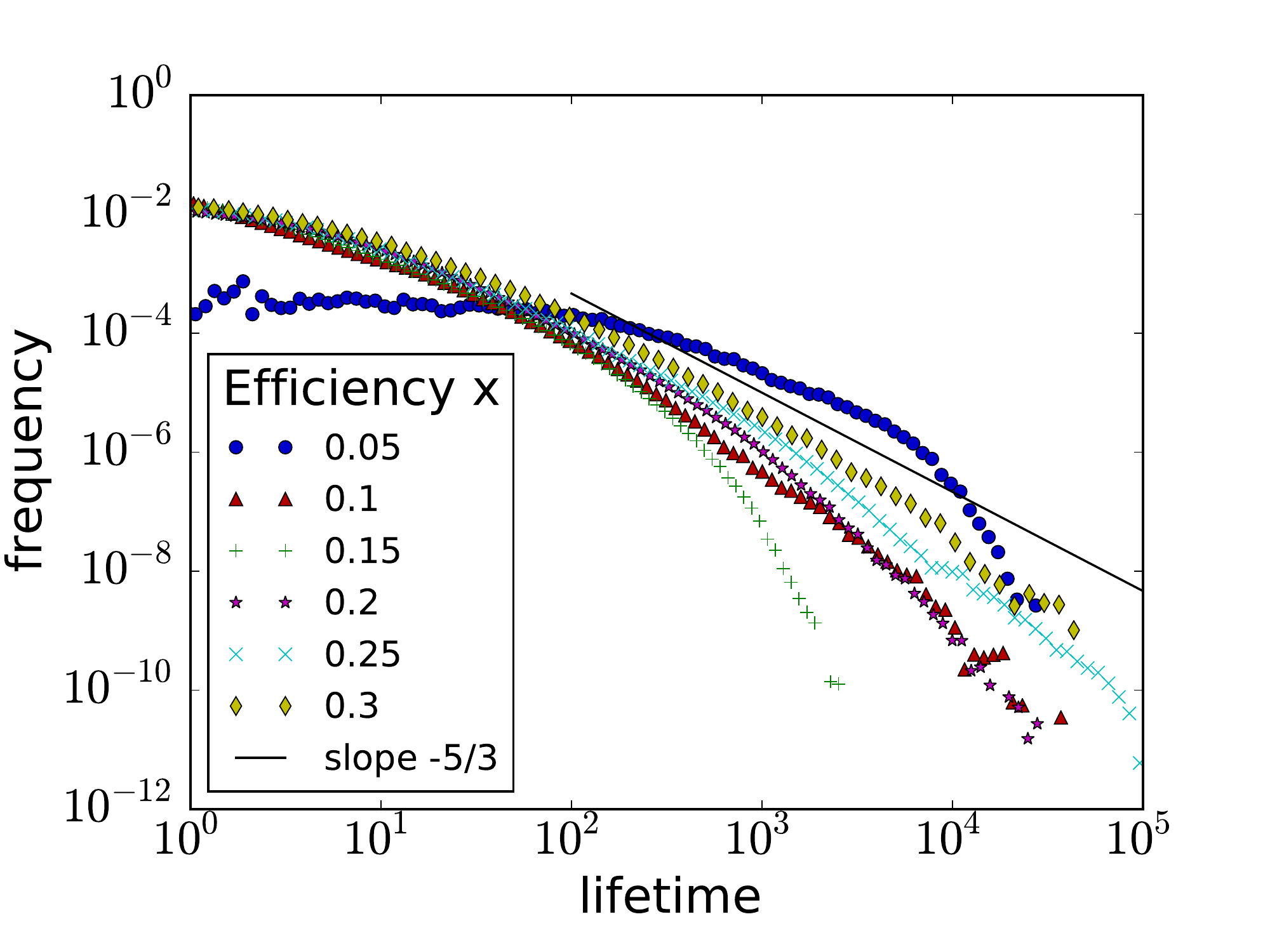}
&
(b)
\includegraphics[width = 0.45\linewidth]{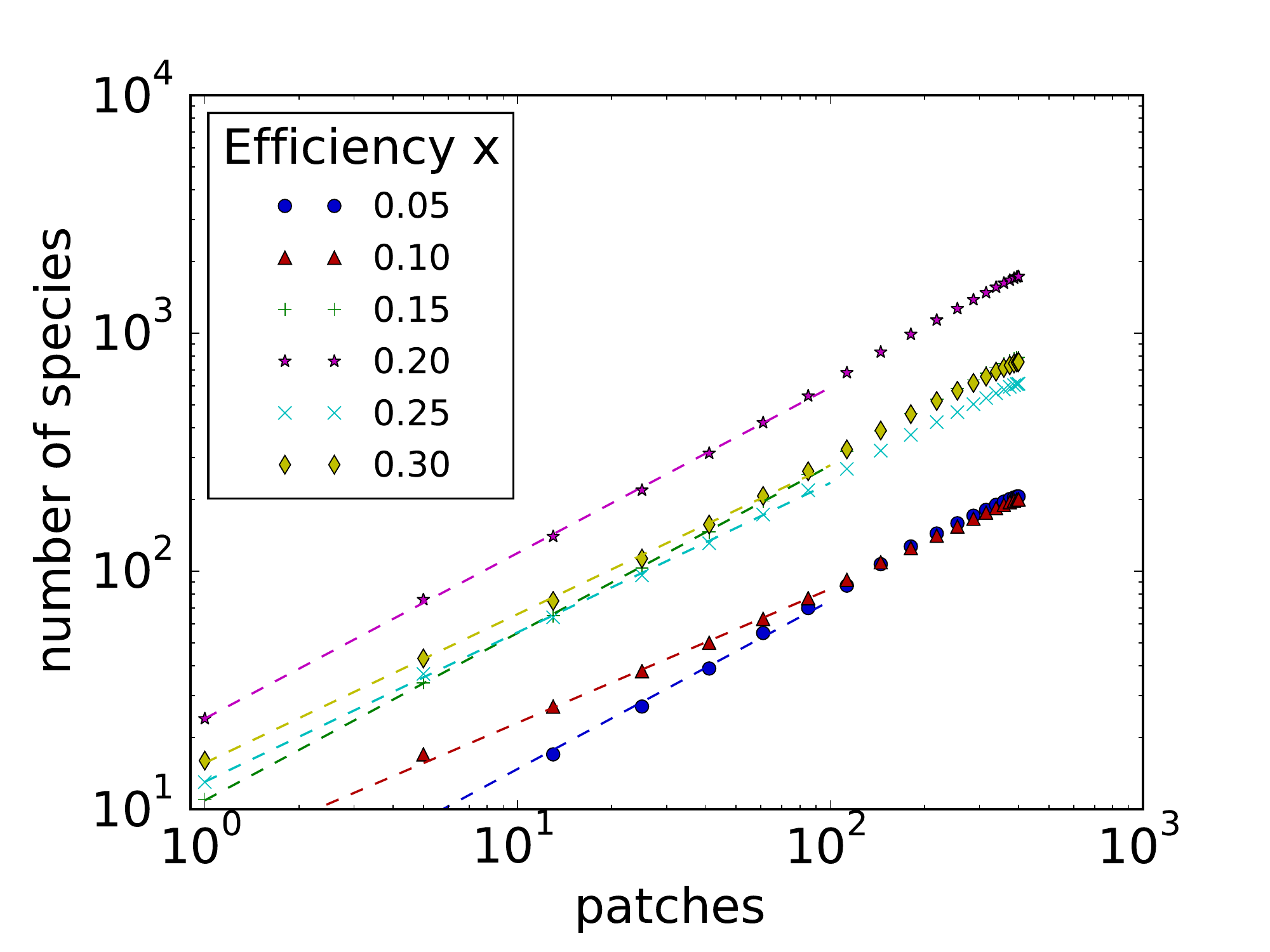}
\\
(c)
\includegraphics[width = 0.45\linewidth]{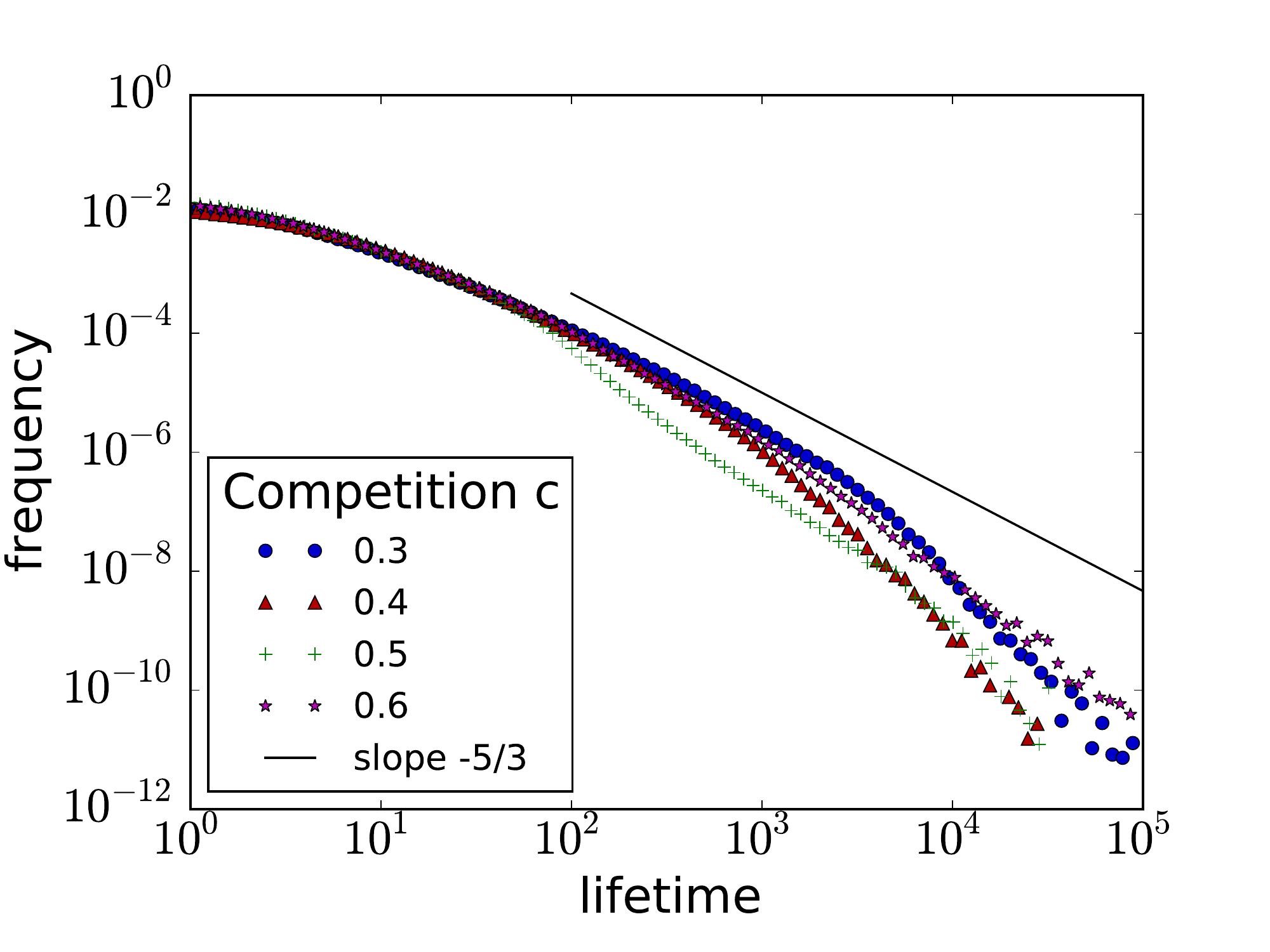}
&
(d)
\includegraphics[width = 0.45\linewidth]{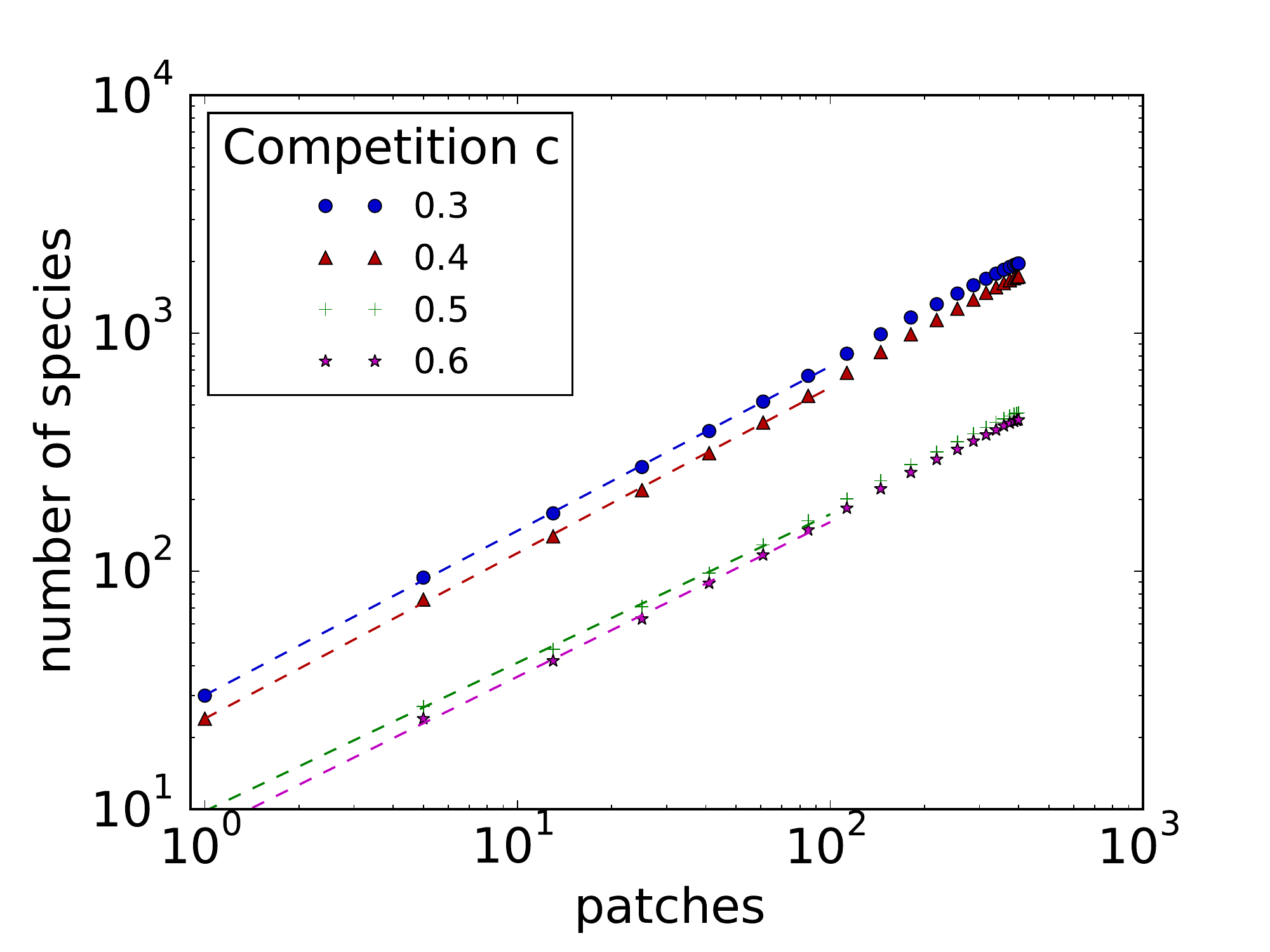}
\\
(e)
\includegraphics[width = 0.45\linewidth]{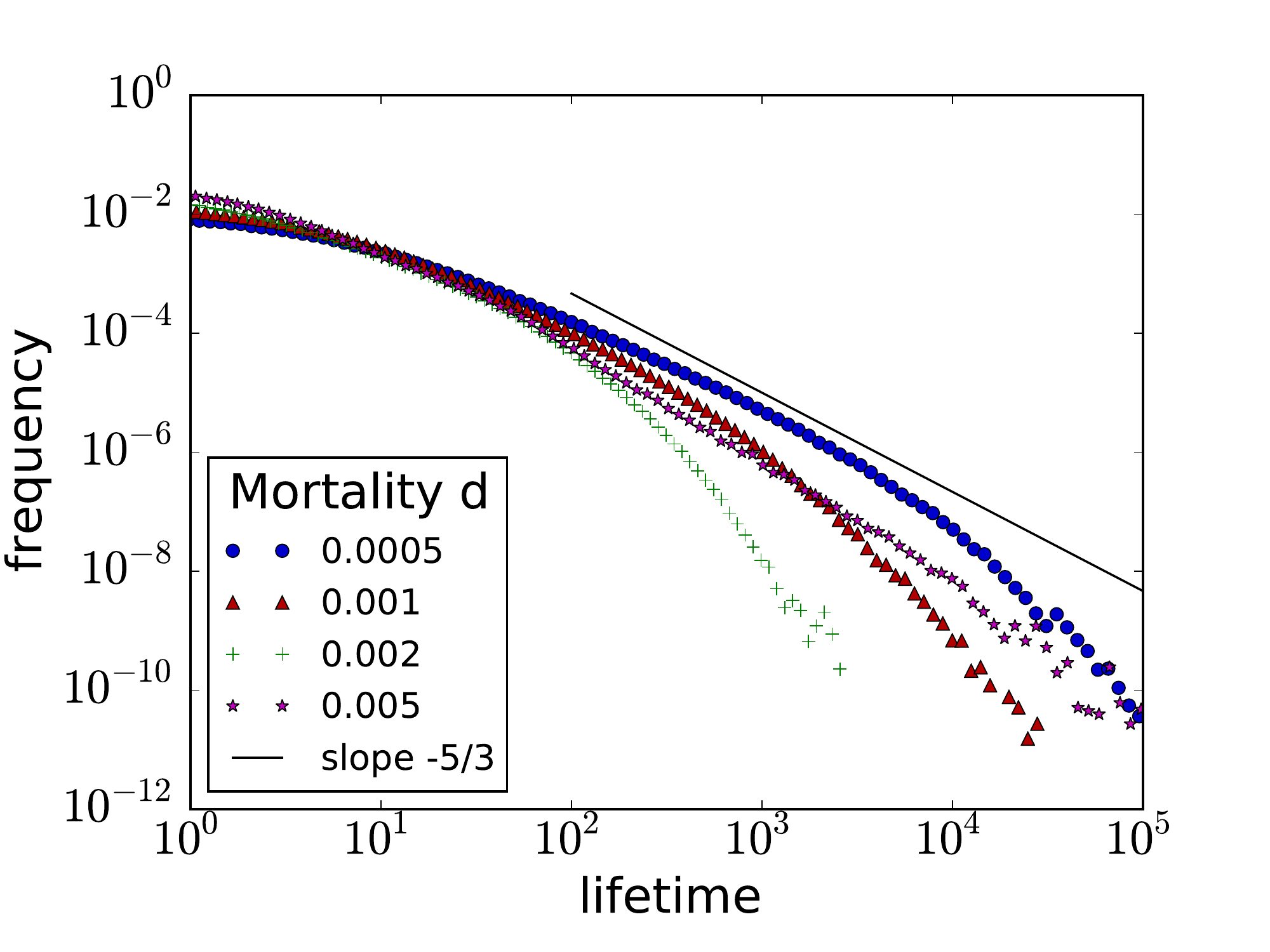}
&
(f)
\includegraphics[width = 0.45\linewidth]{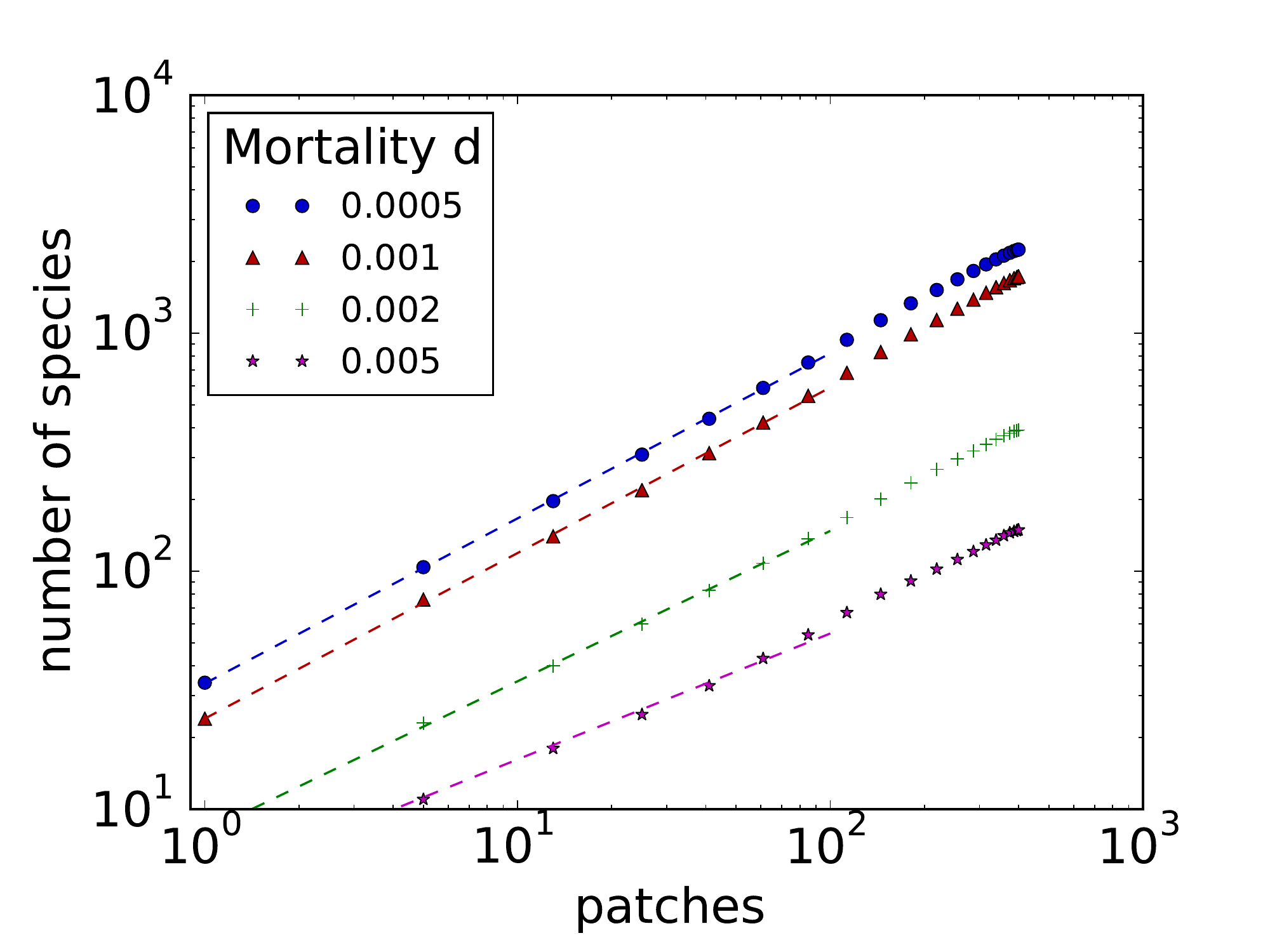}

\end{tabular}
\caption{Influence of parameters x, c and d on lifetime distributions (a,c,e) and SAR (b,d,f). Curves for $x=0.2$, $c=0.4$ and $d=0.001$ represent the standard parameter set. 
The straight line in panels (a,c,e) represents a power law with an exponent of $\alpha = \frac53$, for better comparison. The dashed lines in panels (b,d,f) are power-law fits over the first part of the curves (up to 100 patches), giving the exponents 
0.71, 0.56, 0.70, 0.70, 0.63 and 0.63 for efficiencies 0.05, 0.1, 0.15, 0.2, 0.25 and 0.3 (panel b); 
0.69, 0.70, 0.62 and 0.65 for competition 0.3, 0.4, 0.5 and 0.6 (panel d); and 
0.69, 0.70, 0.63 and 0.53 for mortality 0.0005, 0.001, 0.002 and 0.005 (panel f).}
\label{xcd_variation}
\end{figure}

In agreement with that, we find that harsher conditions (lower efficiencies, stronger competition, higher loss rates) translate into shorter lifetimes. This general trend holds over a considerable parameter range, as shown in Figure \ref{xcd_variation}(a), (c) and (e)). Deviations are only observed, when the emergence of higher trophic levels is no longer possible. In these extreme cases, we observe broader lifetime distributions, possibly reflecting improved survival conditions for species in the first trophic level not experiencing any predation pressure. 

In contrast, we find no clear pattern in how the ecological parameters affect the slopes of species-area relations. Longer lifetimes can translate into flatter SAR (indicating that successful species have more time to invade more patches), but the opposite result can also emerge (possibly indicating that local food webs are saturated, meaning that survival rates are high, but invasion success rates are low).

\bigskip
The simulation presented in Figure \ref{x_timeseries}(e) shows several massive extinction events, where several trophic levels disappear at the same time. These events are usually triggered when a key species on a low trophic level is removed at random (mimicking environmental influences that are not explicitly taken into account), leading to an extinction avalanche that cascades through the food web from bottom to top. We occasionally observed similar extinction cascades also for other parameter combinations but the parameter set in Figure \ref{x_timeseries}(e) seems to produce networks that are particularly susceptible to such events.

We observe that the higher the efficiency, the longer are the emergent food chains. Let us consider those species on an intermediate trophic level. For these species it makes a huge difference whether efficiency is just high enough support few predators in the next trophic level or whether efficiency is so high that the next level is packed with potential predators. In the last case, we can assume that our focal species have a very low survival index and that loosing one single prey might already push their survival indices below 1. Whether or not random extinction events in low trophic levels trigger extinction avalanches thus depends on whether all available niches in the higher trophic levels are occupied or not. 


\clearpage
\subsection*{Variation of evolutionary rules}

\begin{figure}[p]
\centering
\includegraphics[width=0.7\textwidth]{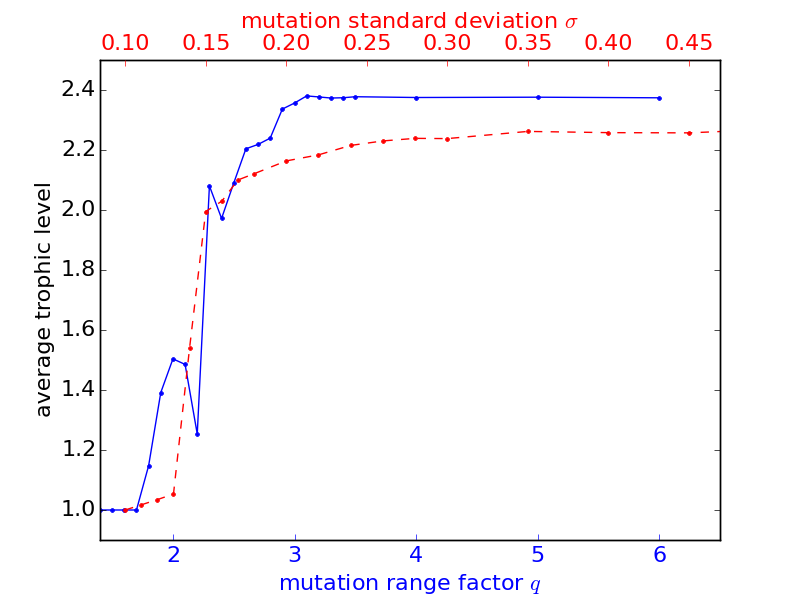}
\caption{Average trophic level of all species within the food web as a function of the mutation range $q$ for the original model version (solid line with scale on the lower x-axis) and the standard deviation $\sigma$ for the modified model version (dashed line with scale on the upper x-axis). All other model parameters are chosen according to our standard parameter set, as indicated in tab. \ref{tab:mut}.}
\label{avg_tl}
\end{figure}

\begin{figure}[p]
\begin{tabular}{cc}
(a)
\includegraphics[width = 0.45\linewidth]{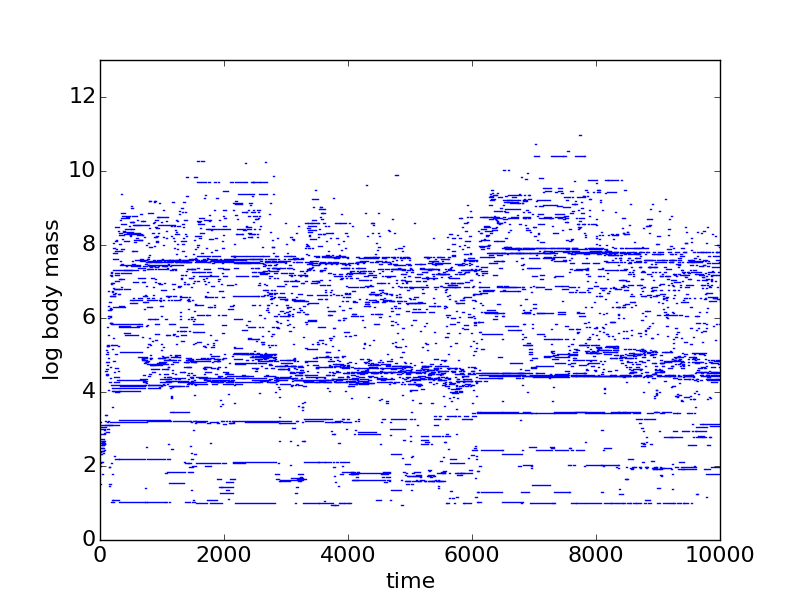}
&
(b)
\includegraphics[width = 0.45\linewidth]{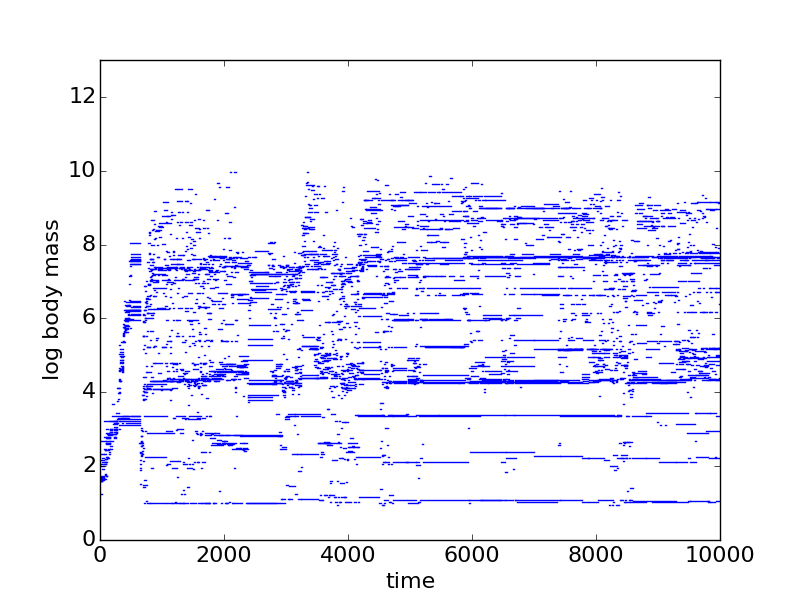}
\\
(c)
\includegraphics[width = 0.45\linewidth]{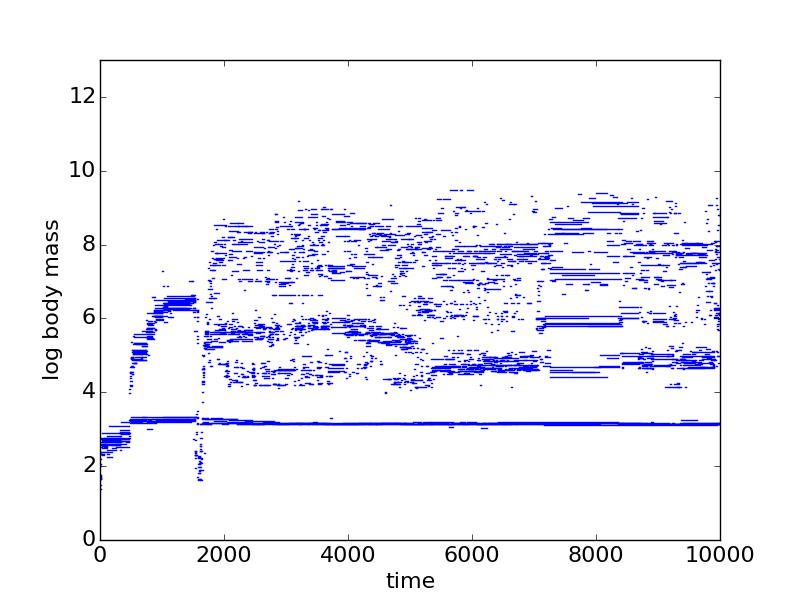}
&
(d)
\includegraphics[width = 0.45\linewidth]{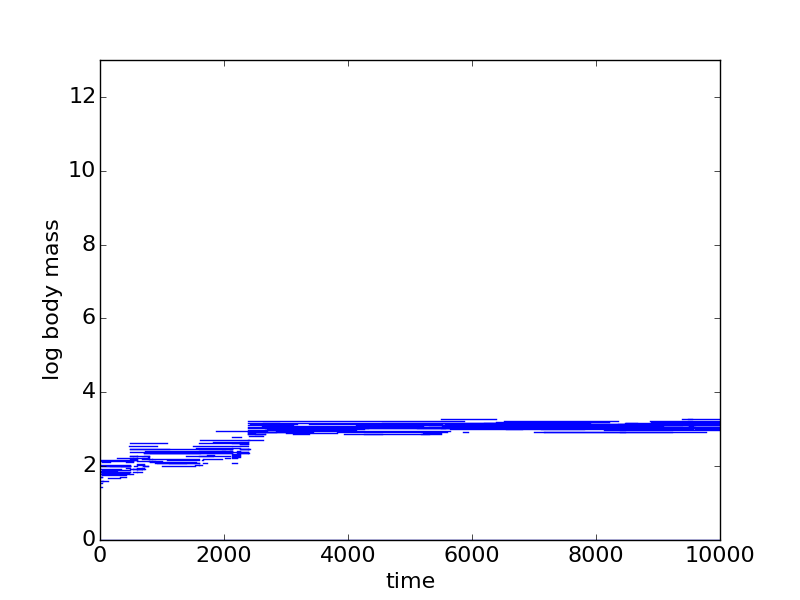}
\\
\end{tabular}
\caption{Evolution of body masses within exemplary simulations runs using a uniform distribution for the mutant body mass and different values for the maximum body mass ratio between a parent and its mutant: (a) $q = 5.0$, (b) $q = 3.0$, (c) $q = 2.6$, (d) $q = 2.0$.}
\label{q_timeseries}
\end{figure}

In a first step, we investigate how variation in $q$ (which describes the maximum body mass ratio between a parent and its mutant species) affects the average trophic level of all species that emerge during a given simulation. This measure serves as a proxy for other structural properties of the resulting food webs. We find that the model is robust against variation in $q$, as long as $q\geq 3$, as shown in Figure \ref{avg_tl}. We furthermore observe a decline of the average trophic level for $q$-values below this threshold, indicating that the emergence of complex, multi-trophic food webs is hampered when mutation steps become too small. Individual time series furthermore reveal that smaller values of $q$ generally lead to slower food web build-up, as shown in Figure \ref{q_timeseries}. 

More precisely, we observe that the emergence of higher trophic levels is suppressed completely, meaning that the food webs contain only species feeding exclusively on the external resource, if $q\leq 1.7$. In this case, species evolve to rather large body masses, so that new mutants are unable to occupy a second trophic level. (To build up a second trophic level, two conditions must be met: First, the competition pressure on a potential prey species within the first trophic level must be small enough to survive an additional predator. Second, the feeding center of the potential predator has to be large enough to feed on other consumers instead of the external resource. This criterion is difficult to fulfill if prey body masses are large and mutation steps small.)

\begin{figure}[ht]
\begin{tabular}{cc}
(a)
\includegraphics[width = 0.45\linewidth]{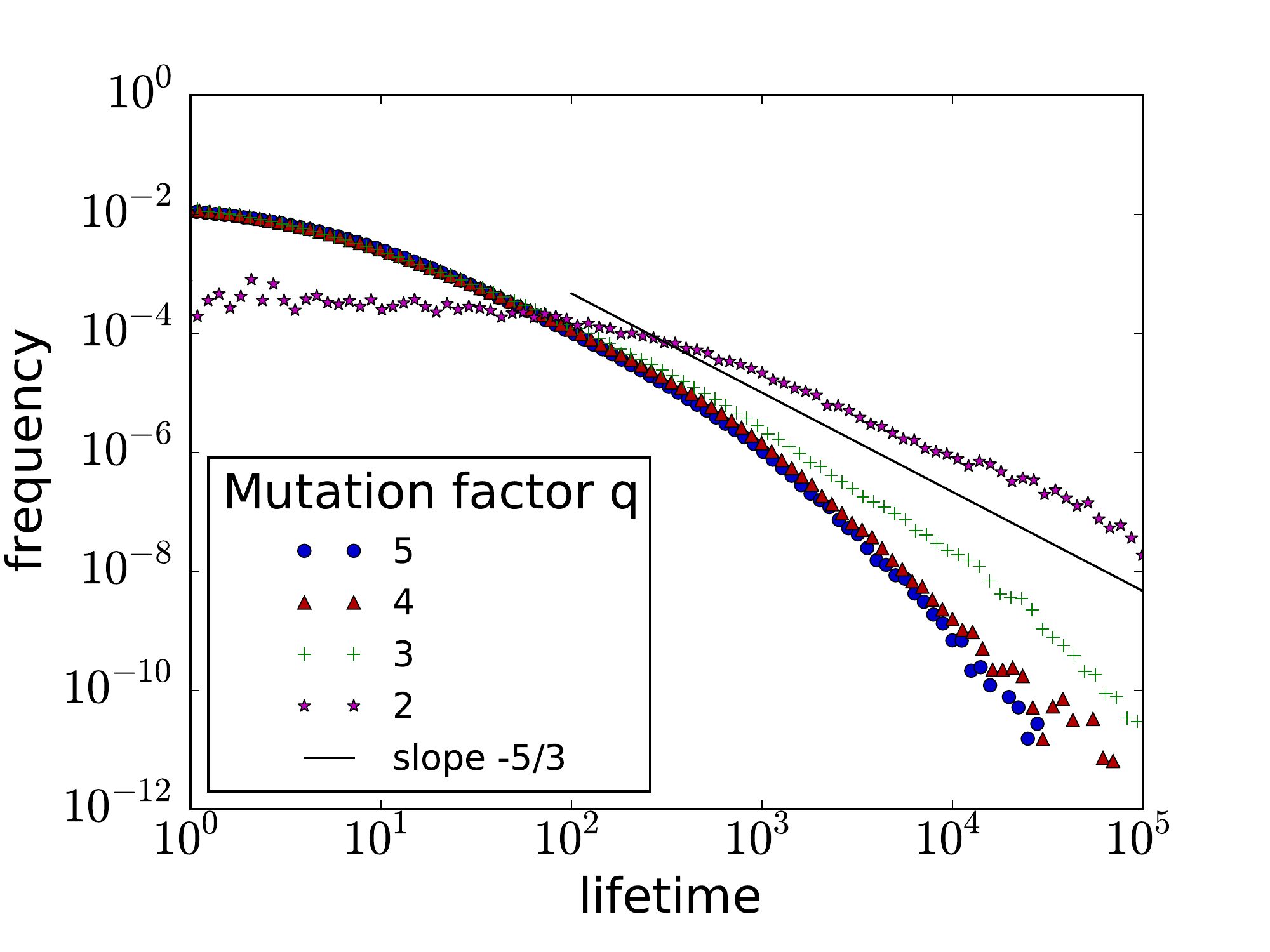}
&
(b)
\includegraphics[width = 0.45\linewidth]{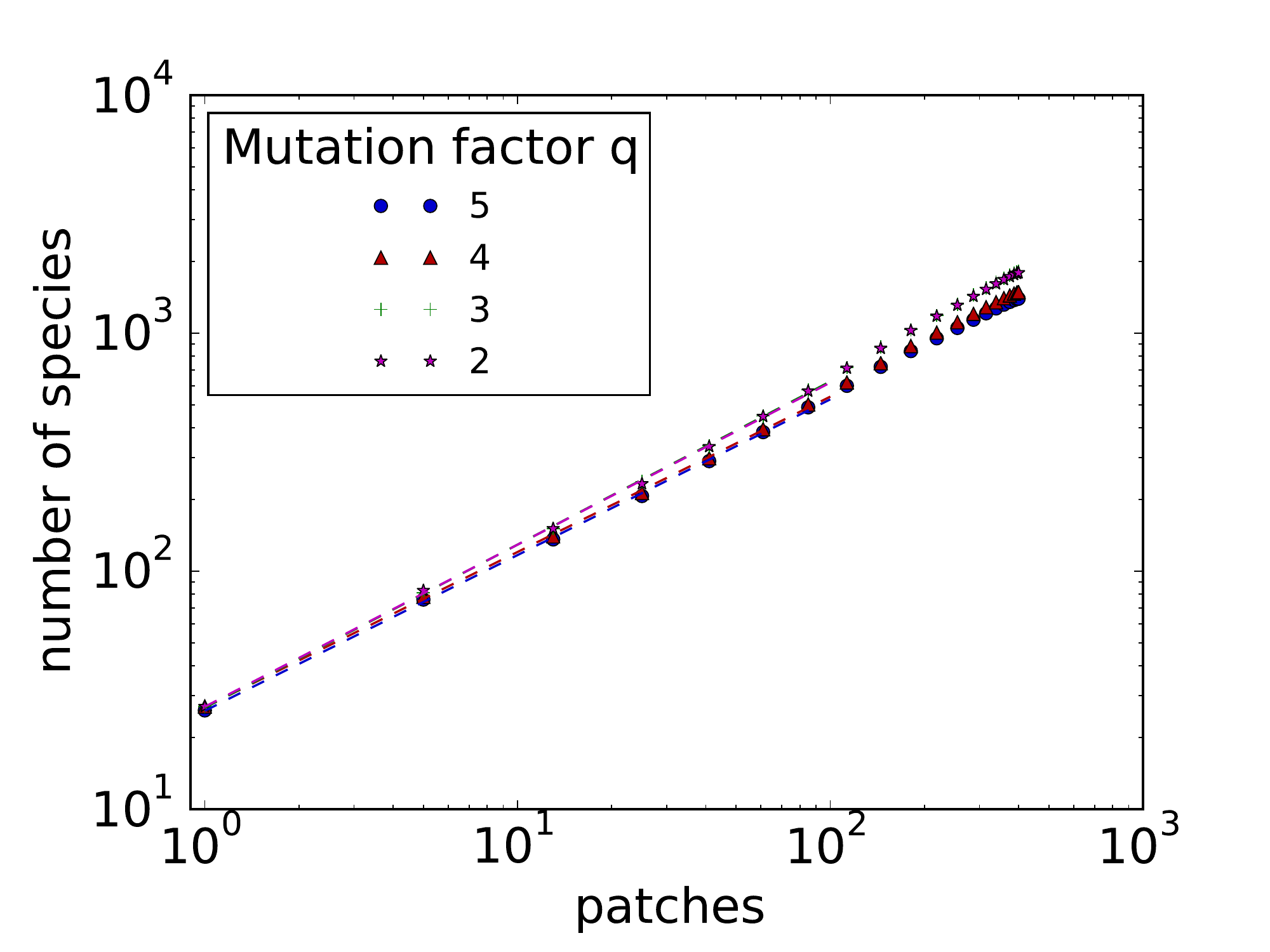}
\\
(c)
\includegraphics[width = 0.45\linewidth]{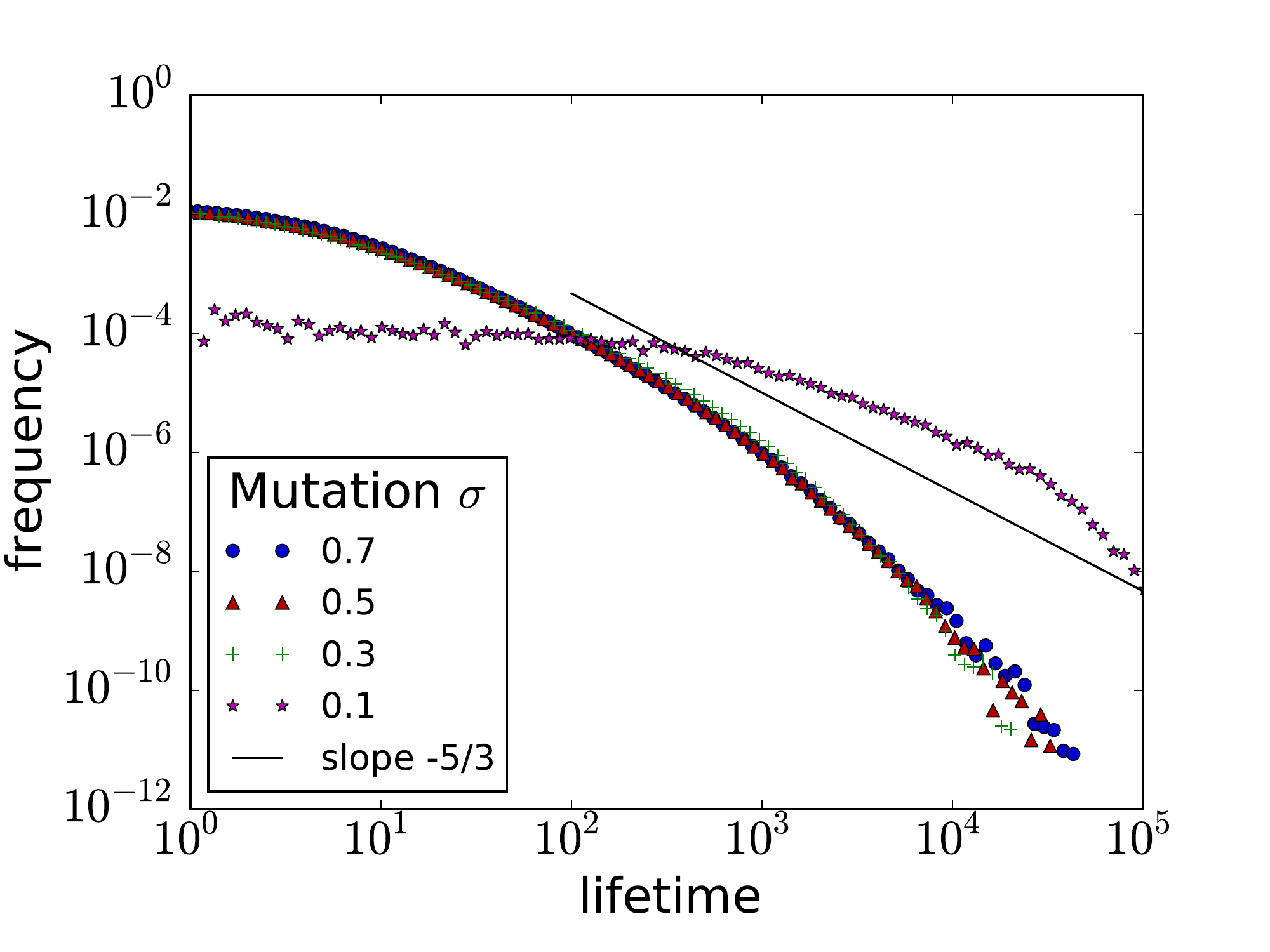}
&
(d)
\includegraphics[width = 0.45\linewidth]{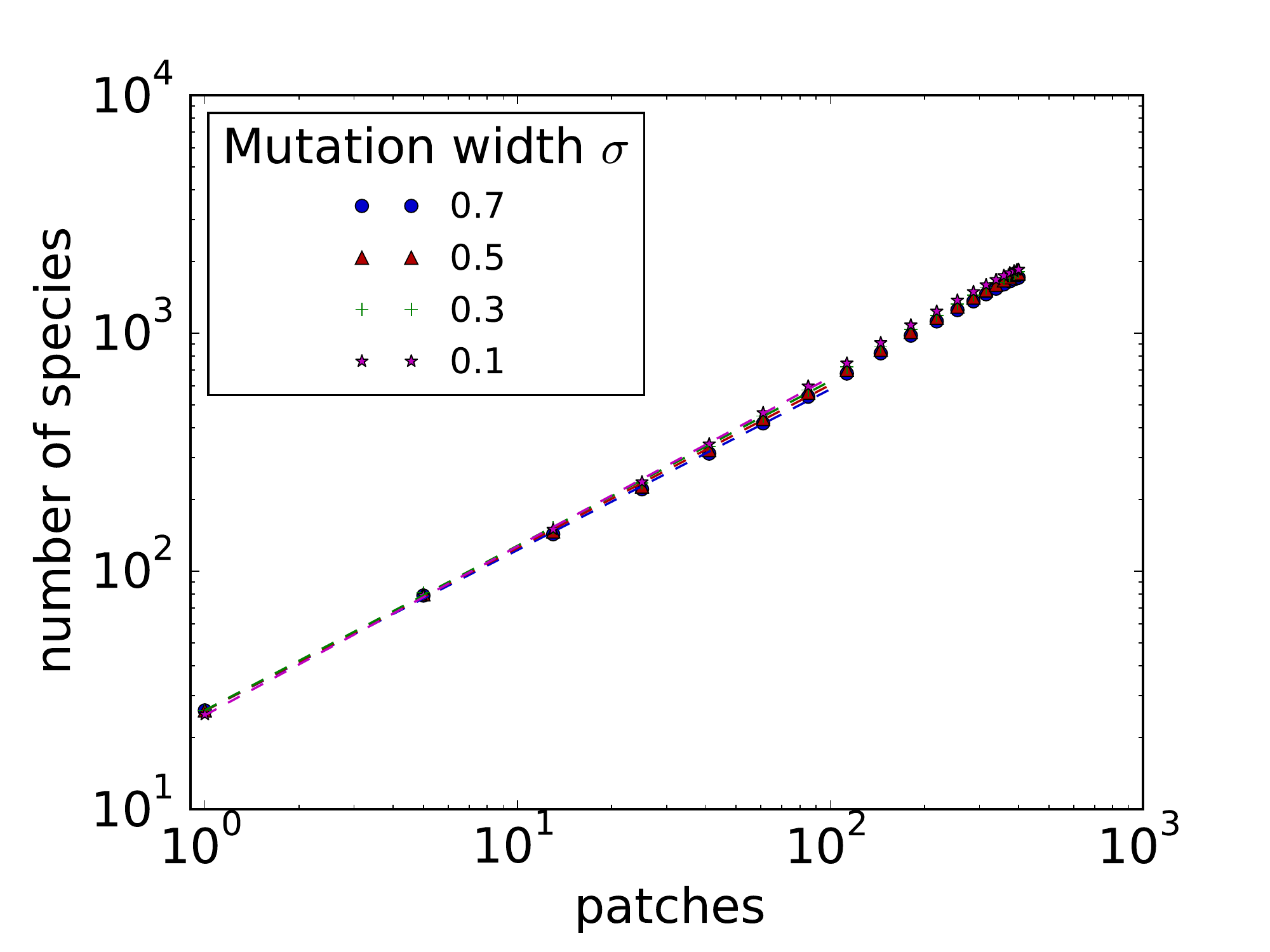}
\\
\end{tabular}
\caption{Lifetime distributions (a,c) and species area relationships (b,d) for different values of the mutation range $q$ for the original model version (a,b) and the standard deviation $\sigma$ for the modified model version (c,d). For the lifetime distributions, all other model parameters were chosen according to our standard set, as indicated in tab. \ref{tab:mut}. The straight line represents a power law with an exponent of $\alpha = \frac53$. To gain results for the species area relationships, we simulated a grid with 20x20 patches and set the migration rate to $\mu_{mig}=10$. The dashed lines are power-law fits over the first part of the curves (up to 100 patches), giving the exponents 0.65, 0.65, 0.69 and 0.68 for mutation factors $q=$ 5, 4, 3 and 2 (b), 0.68, 0.68, 0.69 and 0.71 for mutation widths $\sigma =$ 0.7, 0.5, 0.3 and 0.1 (d).}
\label{robness_q_sig}
\end{figure}

We find that variation in $q$ has very little impact on the resulting lifetime distributions in our model, as shown in Figure \ref{robness_q_sig}(a), if $q$ is chosen large enough to allow for the emergence higher trophic levels. For small values of $q$, we observe a shift towards longer lifetimes, just as in the special cases of very low efficiency, very strong competition or very high loss rates (Figure \ref{xcd_variation}). Again, we hypothesize that this shift reflects improved survival conditions for species in the first trophic level due to the absence of predation pressure. In this case (and in contrast to the results reported in our study!), we find that longer lifetimes do not translate into flatter SAR, as shown in Figure \ref{robness_q_sig}(b). The first trophic level on each patch is simply filled-up until all niches are occupied. The number of locally co-existing species is then determined only by the overall competition strength in the system and species turnover is mostly triggered via random extinction events. The communities within neighboring patches are therefore very similar. Competition between local and invader species results in low invasion success rates and therefore restricted species ranges. 

\newpage
In a second step, we study a more realistic (but also more complex) variation of our model, where the mutant body mass is chosen from a Gaussian (instead of a uniform) distribution centered around the parent body mass. This describes a situation where most mutation steps are rather small (meaning very similar parent and mutant body masses), and were bigger mutation steps occur rarely. For this model variant, we vary the standard deviation $\sigma$ of the Gaussian distribution instead of the mutation range $q$. Both parameters $\sigma$ and $q$ can be easily compared using the standard deviation of an equal distribution, which gives $\sigma_q = \frac{2 \cdot log(q)}{\sqrt{12}}$. 

\begin{figure}[p]
\begin{tabular}{cc}
(a)
\includegraphics[width = 0.45\linewidth]{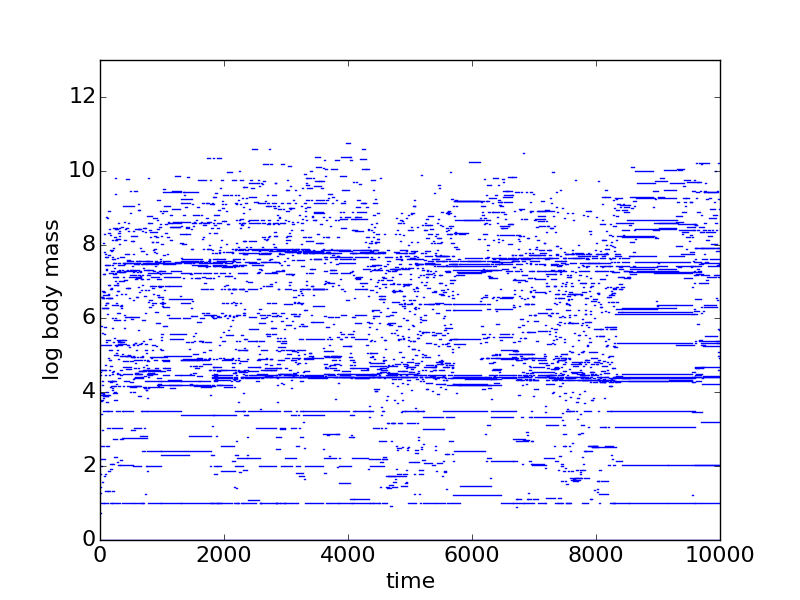}
&
(b)
\includegraphics[width = 0.45\linewidth]{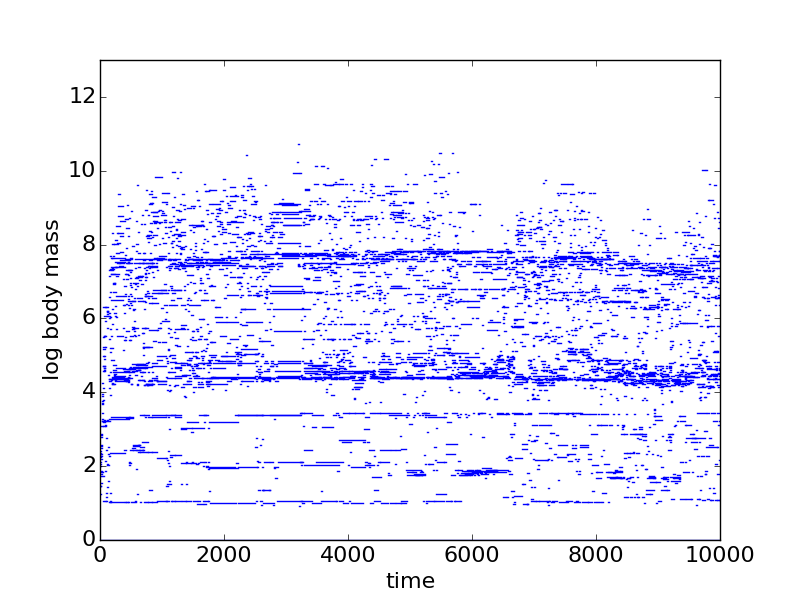}
\\
(c)
\includegraphics[width = 0.45\linewidth]{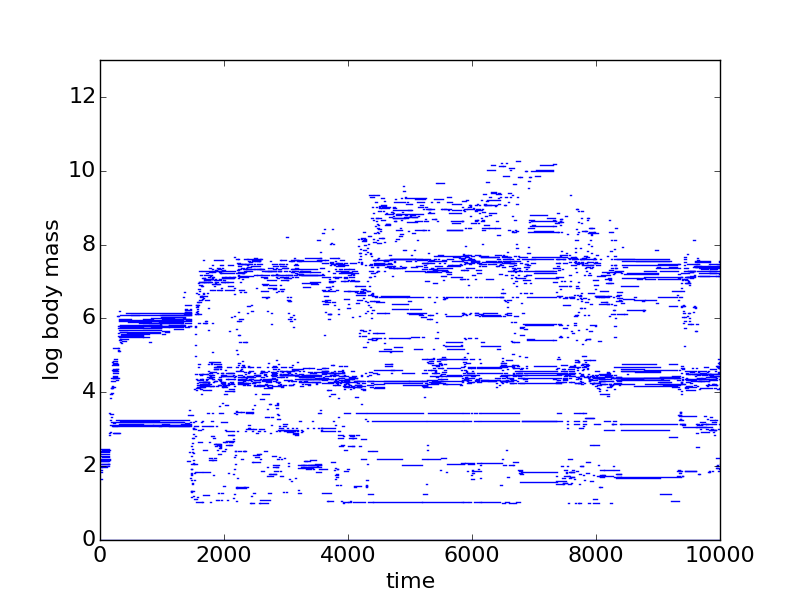}
&
(d)
\includegraphics[width = 0.45\linewidth]{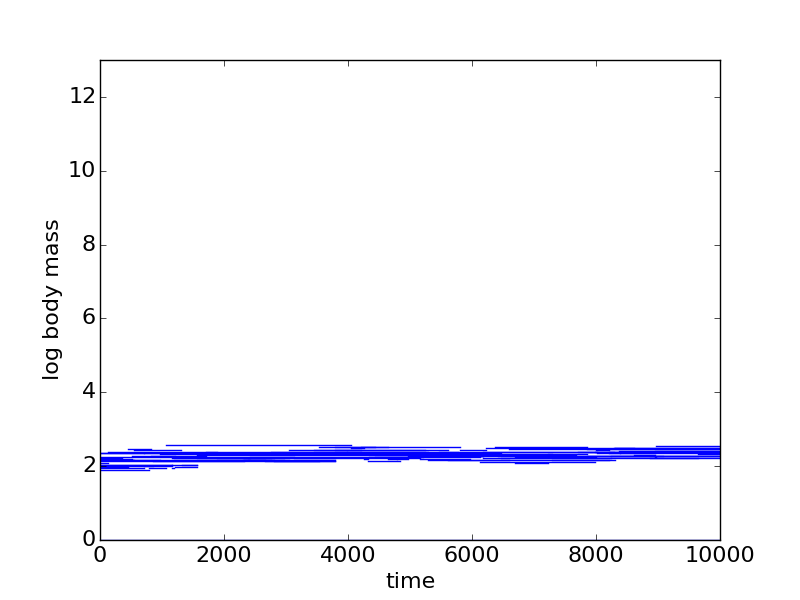}
\\
\end{tabular}
\caption{Evolution of body masses within exemplary simulations runs using a Gaussian distributed mutant body mass and different values for the standard deviation: (a) $\sigma = 0.7$, (b) $\sigma = 0.4$, (c) $\sigma = 0.2$, (d) $\sigma = 0.1$.}
\label{sig_timeseries}
\end{figure}

Using this model modification, we obtain very similar results as discussed above for the original model version, as summarized in Figure \ref{avg_tl}, \ref{robness_q_sig}(c) and (d), and \ref{sig_timeseries}. More precisely, we find again that (i) the average trophic level is mostly independent of $\sigma$ in case $\sigma$ is chosen large enough, (ii) smaller values of $\sigma$ generally lead to slower network build-up and (iii) no higher trophic levels emerge in case $\sigma$ is too small. In summary, we see no remarkable differences between the original model version and the model modification. We therefore use the simpler version in our article in an attempt to minimize model complexity and computational costs.

\clearpage{}

\setcounter{page}{28} 
 




\bibliographystyle{spmpsci}
\bibliography{References.bib}


\end{document}